\documentclass[11pt,fancychapters]{report}
\usepackage[a4paper, total={6in, 8in}]{geometry}
\usepackage{listings}
\usepackage{color}
\usepackage{xcolor}
\usepackage{setspace}
\usepackage{acro}
\usepackage{amsmath}
\usepackage{amsthm}
\usepackage{graphicx}
\usepackage{geometry}
\usepackage{subcaption}
\usepackage{cancel}
\usepackage{tikz}
\usepackage{color}
\usepackage{dsfont}
\usepackage[frozencache,cachedir=.]{minted}
\usepackage{caption}
\usepackage{comment}
\usepackage{glossaries}
\usepackage{authblk}
\usepackage[backend=bibtex]{biblatex}
\usepackage{doi}
\usepackage{colortbl}
\usepackage{booktabs}
\usepackage{hyperref}
\usepackage{multirow}
\usepackage{longtable}
\usepackage{pythontex}
\usepackage{listings}
\usepackage{amssymb}
\usetikzlibrary{shapes.geometric, arrows}

\tikzstyle{startstop} = [rectangle, rounded corners, minimum width=3cm, minimum height=1cm,text centered, draw=black, fill=red!30]
\tikzstyle{process} = [rectangle, minimum width=3cm, minimum height=1cm,text centered, draw=black, fill=gray!30]
\tikzstyle{blocksignal} = [rectangle, minimum width=3cm, minimum height=1cm,text centered, draw=black, fill=blue!30]
\tikzstyle{decision} = [diamond, minimum width=3cm, minimum height=1cm, text centered, draw=black, fill=green!30]
\tikzstyle{textbox}  = [rectangle, rounded corners, minimum width=2cm, minimum height=1cm, text centered, draw=white]
\tikzstyle{arrow} = [thick,->,>=stealth]
\tikzstyle{dasharrow} = [dashed,->,>=stealth]
\tikzstyle{function} = [circle, radius=5pt, draw=black, fill=green!30, text centered]

\newcounter{example}[section]
\newenvironment{example}[1][]{\refstepcounter{example}\par\medskip
   \noindent \textbf{Example~\theexample. #1} \rmfamily}{\medskip}

\newtheorem{theorem}{Theorem}[section]   
\newtheorem{exercise}[theorem]{Exercise}

\theoremstyle{definition}
\newtheorem{definition}{Definition}[section]

\usetikzlibrary{calc,trees,positioning,arrows,chains,shapes.geometric,%
    decorations.pathreplacing,decorations.pathmorphing,shapes,%
    matrix,shapes.symbols}
\DeclareAcronym{etf}{
	short = ETF,
    long = Exchange-Traded Fund,
    tag = abbrev
}

\DeclareAcronym{aum}{
	short = AUM,
    long = Assets Under Management,
    tag = abbrev
}

\DeclareAcronym{sr}{
	short = SR,
    long = Sharpe Ratio,
    tag = abbrev
}

\DeclareAcronym{nyse}{
	short = NYSE,
    long = New York Stock Exchange,
    tag = abbrev
}

\DeclareAcronym{hft}{
	short = HFT,
    long = High Frequency Trading,
    tag = abbrev
}

\DeclareAcronym{pv}{
	short = PV,
    long = present value,
    tag = abbrev
}

\DeclareAcronym{fv}{
	short = FV,
    long = future value,
    tag = abbrev
}

\DeclareAcronym{ir}{
	short = IR,
    long = interest rate,
    tag = abbrev
}

\DeclareAcronym{capm}{
	short = CAPM,
    long = Capital Assets Pricing Model,
    tag = abbrev
}

\DeclareAcronym{apt}{
	short = APT,
    long = Arbitrage Pricing Theory,
    tag = abbrev
}

\DeclareAcronym{sma}{
	short = SMA,
    long = simple moving average,
    tag = abbrev
}

\DeclareAcronym{mvo}{
	short = MVO,
    long = mean variance optimization,
    tag = abbrev
}

\DeclareAcronym{rmse}{
	short = RMSE,
    long = root mean square error,
    tag = abbrev
}

\DeclareAcronym{kNN}{
	short = kNN,
    long = k-nearest neighbor,
    tag = abbrev
}
\hypersetup{
    colorlinks,
    citecolor=black,
    filecolor=black,
    linkcolor=black,
    urlcolor=black
}

\definecolor{DarkerGreen}{RGB}{0,179,45}

\definecolor{Code}{rgb}{0,0,0}
\definecolor{Decorators}{rgb}{0.5,0.5,0.5}
\definecolor{Numbers}{rgb}{0.5,0,0}
\definecolor{MatchingBrackets}{rgb}{0.25,0.5,0.5}
\definecolor{Keywords}{rgb}{0,0,1}
\definecolor{self}{rgb}{0,0,0}
\definecolor{Strings}{rgb}{0,0.63,0}
\definecolor{Comments}{rgb}{0,0.63,1}
\definecolor{Backquotes}{rgb}{0,0,0}
\definecolor{Classname}{rgb}{0,0,0}
\definecolor{FunctionName}{rgb}{0,0,.7}
\definecolor{Operators}{rgb}{0,0,0}
\definecolor{Background}{rgb}{0.98,0.98,0.98}

\lstdefinestyle{python}{
  numbers=left,
  numberstyle=\footnotesize,
  numbersep=1em,
  xleftmargin=1em,
  framextopmargin=2em,
  framexbottommargin=2em,
  showspaces=false,
  showtabs=false,
  showstringspaces=false,
  frame=l,
  tabsize=4,
  basicstyle=\ttfamily\small\setstretch{1},
  backgroundcolor=\color{Background},
  language=Python,
  commentstyle=\color{Comments}\slshape,
  stringstyle=\color{Strings},
  morecomment=[s][\color{Strings}]{"""}{"""},
  morecomment=[s][\color{Strings}]{'''}{'''},
  morekeywords={import,from,class,def,for,while,if,is,in,elif,else,not,and,or,print,break,continue,return,True,False,None,access,as,,del,except,exec,finally,global,import,lambda,pass,print,raise,try,assert},
  keywordstyle={\color{Keywords}\bfseries},
  morekeywords={[2]@invariant},
  keywordstyle={[2]\color{Decorators}\slshape},
  emph={self},
  emphstyle={\color{self}\slshape},
  breaklines=true
}

\newtheorem{exmp}{Example}[section]

\newcommand{\splitatcommas}[1]{\begingroup\lccode`~=`, \lowercase{\endgroup
    \edef~{\mathchar\the\mathcode`, \penalty0 \noexpand\hspace{0pt plus 1em}}%
  }\mathcode`,="8000 #1%
  }

\tikzset{
>=stealth',
  punktchain/.style={
    rectangle, 
    rounded corners, 
    draw=black, very thick,
    text width=10em, 
    minimum height=3em, 
    text centered, 
    on chain},
  line/.style={draw, thick, <-},
  element/.style={
    tape,
    top color=white,
    bottom color=blue!50!black!60!,
    minimum width=8em,
    draw=blue!40!black!90, very thick,
    text width=10em, 
    minimum height=3.5em, 
    text centered, 
    on chain},
  every join/.style={->, thick,shorten >=1pt},
  decoration={brace},
  tuborg/.style={decorate},
  tubnode/.style={midway, right=2pt},
}

\makeglossaries

\newglossaryentry{non parametric}
{
    name=non parametric,
    description={Non parametric statistics does not assume that the underlying probability distribution has a predefined form, e.g., of type normal or exponential, and nevertheless is able to apply statistical inference to the problem. }
}

\newglossaryentry{control interval}
{
    name=control interval,
    description={For a random variable $X$, a control interval is a segment $[low, high]$ such that the probability of having a random variable materialising outside the segment is low}
}

\newglossaryentry{ML model}
{
    name=ML model,
    description={Depending on the learning task this may stand for different things.  In general a ML model some determining function that has predictive capability.   For example, in the case of a clarification ML task it will be a function that given a new objected, e.g., en image, 
determine its type, e.g., whether it is a cat or a dog. }
}

\newglossaryentry{bootstrapping}
{
    name=bootstrapping,
    description={Given a sample, $P$, from some distribution $F$, the empirical distribution of the sample $S$ represents the distribution, $P$, if the sample is big enough.  We can thus re-sample from $S$ with replacement to obtain fresh samples that approximate fresh samples from $P$.  This process is called bootstrapping.   }
}

\bibliography{main}
\begin{document}

\title{%
A Practical Approach to Combinatorial Test Design
}

\author[1]{Eitan Farchi}
\author[2]{Debbie Furman}
\affil[1]{IBM Haifa Research Lab, farchi@il.ibm.com}
\affil[2]{IBM Infrastructure, debbyt@us.ibm.com}
\date{2024}

\maketitle
\pagenumbering{gobble}
\newpage
\pagenumbering{roman}
\tableofcontents
\newpage
\pagenumbering{arabic}

\begin{abstract}
Typical software has a huge input space. The number of inputs may be astronomical or even infinite. Thus, the task of validating that the software is correct seems hopeless. To deal with this difficult task, Combinatorial Test Design (CTD) can be used to provide reduction of the testing space and high quality and efficient testing.  The application of CTD is largely determined by the quality of the CTD model. This book covers the CTD test design methodology and CTD modeling in details. It elaborates on the process of constraints definition.  It also explains how to best define your coverage requirements to direct and focus your tests.  It is hard to create good CTD models without a good grasp of the implementation of CTD tooling. To that hand, the book also takes a deeper dive into covering principles and algorithms needed to build CTD tooling. Hands on exercises are used throughout the text and help create a clear understanding of the concepts covered within this book.
\end{abstract}

\chapter{Introduction}

Typical software has a huge input space.  The number of inputs may be astronomical or even infinite.  Thus, the task of validating that the software is correct seems hopeless.  To deal with this impossible task different approaches are practiced.  They all attempt to focus on parts of the inputs that matter.  

One approach includes the use of user profile that captures the more frequent use cases as anticipated while the software is developed or observed as the software is being deployed.  This approach is sensitive to errors in the estimation of the usage profile and changes over time to the profile. Another approach considers dividing the inputs into subsets that are expected to either exhibit a problem or not exhibit a problem and then tests a representative from each input subset.   Such approaches are referred to as the application of a coverage methodology.  Even when coverage approaches are applied we can still end up with a representative number from each subset of the input space that is too large.  

Consider the following example.  We need to test a code comprised of consecutive if-else statements. Thus we have say 10 consecutive statements that looks like \\
$if(condition) then "do somthing" else "do somthing else"$. The number of paths through this code snippet is $2^{10} = 1024$ which is too big to implement if you adapt the approach of requiring the coverage of all paths through the code. More importantly it grows exponentially as the number of consecutive if-else statements is increased.  To tackle that further reduction is required.

Combinatorial Test Design (CTD) is one such way to provide further reduction of the testing space.  The National Institute of Science and Technology  (see \href{https://www.nist.gov/programs-projects/combinatorial-testing}{link}) gives the following explanation of the CTD approach to test reduction - 
"Combinatorial or t-way testing is a proven method for more effective testing at lower cost. The key insight underlying its effectiveness resulted from a series of studies by NIST from 1999 to 2004. NIST research showed that most software bugs and failures are caused by one or two parameters, with progressively fewer by three or more, which means that combinatorial testing can provide more efficient fault detection than conventional methods. Multiple studies have shown fault detection equal to exhaustive testing with a 20X to 700X reduction in test set size.  New algorithms compressing combinations into a small number of tests have made this method practical for industrial use, providing better testing at lower cost. "

\chapter{A roller coaster introduction  to Combinatorial Test Design }
\section{Introduction}

This chapter describes how to effectively design tests using Combinatorial Test Design (CTD).  We  assume a test objective has been identified and the strategy or general scenarios that will help meet the test objective defined.   For example, the test objective may be to determine data is not lost on power failure when memory is written.  The test strategy or scenarios may constitute sequences of writes of different sizes to memory occurring in parallel to a power failure.  After recovery from the power failure, memory is checked to determine if it correctly contains the data that was written before the failure occurred.

Typically, once the test objective and scenarios or strategy that meet them are identified, a more detailed step by step description of the tests is given and variables and/or input variables identified. In our running example the variables constitute of types of power failure, number of writes until the power failure, types of writes, whether cache is turn on or not, and the amount of data written on each write.   The test may be defined as follows:

\begin{itemize}
\item input: type of power failure
\item input: length of write sequence of type short, medium or long
\item input: cache is on or not
\item Randomly choose the number of writes k in a range of write lengths that was defined as short, medium or long 
\item k times
\begin{itemize}
\item randomly choose the size of the write, and execute the write operation
\end{itemize}
\item Cause a power failure of the type specified in the input
\item Recover from the failure and check the memory state to determine if data was lost
\end{itemize}

Assuming there are 4 types of power failures and we are interested in short, long and medium sequences of writes, the above test has $4*3*2 = 24$ combinations.  In the above design we assumed that these are the variables we would like to cover their interaction and are concerned that their interaction will cause a problem to materialize.  As we will shortly see CTD will help us do that.  In contrast, we assumed that the amount of data written on each write is not of a concern.  Thus, we just randomized the amount of data written on each run.   This will be further explain in \ref{choose} below.  

The rest of this chapter explains the additional steps that should be taken when applying CTD.  It can also be used as a template or a check list.  Each of the following sections should be considered and realized.  Eventually a CTD test design document should point to the Cartesian product model(s) that are used in the design.  The following sections also specify when this should occur.

\section{Variables definition}
\label{choose}
  List the variables used in your test.  In our case they are the power failure type, cache state (on or off), number of writes before the power failure occurred, and the amount of data written.        

\subsection{Decide which variables are part of the model and which should be randomized}

The first thing to determine is which variables and variable interactions we are concerned about, i.e., expect that their interaction may cause a problem to materialize.  For the sake of this example, we decide that the interaction of power failure type, cache state, number of writes before the power failure occurred may cause a problem to materialize.  In contrast, we don't believe that the size of the write and its interaction with other values will cause a problem to materialize.   The best practice in this case is to make the first three variables part of the CTD model and randomize on the write size at test execution time; this is actually helping us check our assumption that the size is immaterial to the forcing of a bug. 

\subsection{Sub-domain partition variables values}

Next we need to determine the values of the variables.   Some of the values are easy.   For example, the cache could be on or off.  The types of power failures are just a simple list.  The number of writes is a challenge - it's a theoretical infinite set of possibilities...We handle such cases by applying sub-domain partition.  We are concerned about very short writes sequences and very long write sequences.  For completeness, medium length sequences of writes are also needed.  Thus, we divide the length of the sequence into three sub-domains denoted as small, medium, and long.   Each sub-domain is given concrete values.  Thus, $ 1 \leq small < 10 $, $ 10 \leq medium \leq 100$, $100 < long $.  The best practice will be to randomly chose a concrete value in these ranges for each execution.   Thus, this is handled in two phases:

\begin{enumerate}
\item An (abstract) value is chosen at the CTD model level. e.g., the number of writes are chosen to be small.
\item A concrete value is chosen to be used when executing the test.  For example, a length between 1 and 10 is chosen at random to represent the abstract value small.
\end{enumerate}

The next three steps captured in the sections below are typically captured/obtained in/from the CTD tool being used.  The test plan document should probably point to the tool artifacts/models/reports at this point. 

\section{Define constraints}

Some combinations may be impossible.  We want to capture that by defining some logical conditions.  For example it may be the case that some types of power failures can not happen if the cache is on.  This subsets the free combinations of variable values resulting in a subset of combinations that can happen.

\section{Define the level of values interaction}
\label{coverage}

Based on statistics it's well known that most of the defects are found by exercising combination of two values.  Exercising all possible two values combinations is called 2-way coverage.  In addition, almost all defects are found by exercising all combinations of 3 values.   Thus, the best practice is to default to 2-way coverage and add combinations of three values that we want to cover based on some test concern.   This customizes the interaction of values we are interested in testing based on our test concerns.   

\section{Obtain an optimized plan}

Next, a CTD tool is used to produce as minimal or close to minimal set of possible combinations of the entire set of variables that covers the required 2 values and three values interaction\footnote{Typically for a realistic model you will have to use a tool to produce the optimal test plan as the number of combinations is too large. }.       

\section{An environment model}
\label{env}

In the previous sections we have covered the steps needed to create a scenario based model.  Another model that arises in practice is the environment model.  An environment model represents different states we would like to cover.   Typically we would like, in principal, to run a set of scenarios in all of the states represented by the environment model. As that may result in too many runs, we optimize by applying CTD to the environment model as well.  We can also combine an environment model with a scenario model and apply CTD to the joint model.   This results in a stronger test plan.

The creation of an environment model is similar to the creation of a scenario model but instead of starting with a scenario you go straight to the variable definition step (see section \ref{choose} above).  

As an example of an environment model for our running example consider the following variables: hardware type, age of the memory, temperature in which the scenario executes, etc.  Note that the cache variable may also be considered as an environment variable.  It only makes sense to do that if for all the scenarios we are considering cache is a variable of interest\footnote{In fact, the environment model was the first application of CTD when it was first introduced in Bell Labs.} 

\section{Reviewing a model}

Test planning in general and the review of a test plan created using CTD in particular is a very important quality action that increases the effectiveness of the test plan.   Specifically, in the context of the CTD methodology there are specific questions to be asked/things to consider that facilitate the finding of problems in the test plan.   In this section we provide a list of guiding questions with a short motivation to each question.  Two cases are considered below: a case in which a model representing a scenario is being reviewed and a case in which a model representing the environment is being used.     

\subsection{Review of a scenario model}

\begin{enumerate}
\item
What is the scenario we are testing?  A basic understanding of the high level scenario we are testing is needed in order to review the model.   For example, this scenario opens a file for read or write, writes to or reads from the file and then closes it.   As a quick exercise: can you think of the matching model?   The number of possible combinations should be in the thousands... 
\item
The feasibility question. Consider combination of values of this model.  Can they be mapped to concrete executable tests that are instances of the high level scenario?  If the answer is no then we have a model that is not feasible and typically it is not desired\footnote{If the model is not used for testing but in order to review some design work this may be ok.   Models created for the purpose of reviewing design and requirements are currently out side the scope of this document.} 
\item
Can all parameters in the model cause a problem or are there parameters in the model that may not cause problems? Typically we only want parameters in the model whose occurrence and interactions will reveal problems.  Other parameters we are not concerned about: we assume that their occurrence will not cause problems.   The best practice is to randomize on these parameters at the concrete test level in order to better check this assumption.    
\item
Have you done sub-domain partitioning on the values of the model parameters and is it required?   Not going through the sub-domain partition process creates models that are too concrete and plans that are unnecessarily big.
\item
Are all the legal combinations in the Cartesian product a possible test?   This questions checks if the model constraints were captured correctly.   
\end{enumerate}

\subsection{Review of an environment model}

Just change scenario to state and test to state in the list of questions given in the previous subsection...

\section{Cross test cycles optimizations}

In section \ref{coverage} we outlined the best practice used to obtain the desired level of variable's interaction coverage.   Sometimes a test cycle is limited by the number of tests you can run, $n$.   In order to overcome that implement the following procedure\footnote{If the tool does not support the procedure below change it!}:

\begin{enumerate}
\item
Require at most $n$ tests and have the tool obtain as many of the desired interaction combinations as possible
\item
Run the obtained combinations at the first cycle and log which tests passed or failed
\item 
Feed the information of the tests that passed to the tool and require to augment the test with another batch of at most $n$ tests that cover as many of the required missing interactions as possible
\item
Repeat the above procedure until you reach 100$\%$ coverage or you run out of test cycles.
\end{enumerate}

In the next chapter will will take a deeper dive into what it takes to build a model.

\chapter{How to build a model - questions to ask}

As a tester you are given a new item to test.  The new item typically maps to some use case and includes interfaces you will be accessing in order to test the system.  The first thing you need to determine is what is interesting in this item you need to test.  Interesting here means how to trigger the system under test in a way that stands the highest probability of finding a problem.  Another way to think about it is you would like to test the parts of the system under test that represent the highest risk.  For example, you may be given several use cases and several interfaces but you are more concerned with one of the use cases and a particular way of using an interface. Your concern will guide the definition of the domain you are attempting to test.  Typically, the determination is not going to be complete and there will be many ways to test the system that address your concerns.  Thus, determining the test domain based on your concerns will not be sufficient to enable your tests, namely, you will still end up with a huge space of possible test triggers to exercise. 

\begin{example}
We need to test a file system implementation. We are considering three use cases.  One use case retrieves the meta data of a file while another reads from a file and another writes to it.  In addition, we are working with files that are remote or local.   In fact, based on previous experience we are very much concerned with getting the meta data of remote files. This focuses and defines our test domain but we are still left with a huge space of possible tests as the remote files could reside in different parts of the network, there are different types of meta data retrieval and different file  permissions.  
\end{example}

The formal representation of a test model consists of attributes and values along with constraints that are used to help us focus on our concerns but also are utilized to define combinations that are impossible.  As mentioned above, once constraints are defined, the set of possible tests may still be huge.  In chapter \ref{ChapterCoverageReduction} we will deep dive into how to reduce the huge test space to a feasible set of tests to execute. 

Next we will discuss each item introduced in modeling the test space separately. 

\section{Attributes}

Attributes are the points of variation within the testing space.  Attributes can be found in requirements, design documents, use cases, existing tests, problem records or customer requests for enhancements.  Attributes can also be thought of as objects in the system under test.  We can ask a set of questions to help identify what attributes to include in the test model.  

The questions follow two principles.  The first principle is that anything that changed is more sensitive and probably should be focused on for the testing purposes. The second principle is specific to the type of mistakes people make in software development.  It is known that interfaces are a source of miscommunication between teams and are sensitive to issues as a result.  In addition, taking into account all of the environment states in which the software operates is hard and error prone.  We thus end up with the following questions. 

\begin{enumerate}
\label{checkList}
    \item What is being created or altered in the system under test?
    \item What are the states of the system under test?
    \item What services, i.e., interfaces, are provided to interact with the system under test?
    \item Which of the documentation you have identified can help motivate the testing, e.g., requirements document, design document, and existing tests, discuss changes or focuses on interfaces or focuses on the environment?     
\end{enumerate}

\begin{definition}
A documentation in the context of software development is any artifact that is used to describe or explain the software including the software itself.
\end{definition}

Note the \textbf{best practice} is to use any source of documentation that is relevant to the software in order to test the system.   In fact, the more the merrier as we can compare between them.  Any inconsistency between artifacts is value in itself.  It is either a problem in the project documentation or a software issue that was found without running a single line of test code.

\begin{exercise}  \label{Exercise:SourceInfo}
We need to test an implementation of the list abstract data type. A list is a sequence of records and its typical operations include inserting an element, removing an element and searching for an element.   A list may have several implementations, e.g., an array implementation, a pointer based implementation and other implementations.  Can you suggest sources of information that will help you test the list implementation?  Choose the best answer below.
\begin{enumerate}
\item Don't use information on list implementation that you can find on the net as it may be unreliable.
\item Search for an array implementation of a list and use it as  source of information for testing.
\item Search for a pointer based implementation of a list and use it as a source of information for testing.
\item Combine the last two answers and also search for an abstract data type definition of a list and use all of that sources of information to design your test plan. 
\end{enumerate}
See solution \ref{Exercise:SourceInfo:solution}
\end{exercise}

%TBC - can we incorporate LLM discovery in the CTD learning process?

In order to clarify how to use these questions, we give the following example that highlights how possible answers help identify attributes for the system under test.

\begin{example}
We need to test a file system implementation in which we are focusing on manipulation of files in the system.  Possible answers to the above questions, \ref{checkList}, are
\begin{enumerate}
    \item Files are being created or altered in the system under test
    \item Files can exist, not exist, be opened or closed
    \item Services include creating a file, opening a file, reading from or writing to a file, closing a file and deleting a file
    \item 
    \begin{enumerate}
        \item Design may include characteristics including length, special characters or mixed case of the file name
        \item Design may also include the location of a file in a root directory, sub-directory or hidden directory
    \end{enumerate}
    
\end{enumerate}

\end{example}

Attributes can be represented as single variables or categories of variables as shown in table \ref{tab:AttributeExamp}.  The first column represents a category and the second column represents examples of variables within that category.
\begin{table}[H]
    \centering
    \begin{tabular}{|l|l|}
    \hline
    \cellcolor{lightgray}\textbf{Example of Categories}&\cellcolor{lightgray}\textbf{Example of Single Variables} \\
    \hline
         Inputs to a system& User name, password, file name\\
    \hline  
         Current state of a system& Free disk space, messages on a queue\\
    \hline
         Stored information& Name, address \\
    \hline
         Configuration variables& Supported browsers, I/O devices \\
    \hline
         Objects in a system& File, queue, or any data structure type \\
     \hline 
         Actions performed on a system& Update, Restart, Shutdown, Hibernate \\
     \hline
        Role of the system's user& Administrator, unauthorized user \\
     \hline
    \end{tabular}
    \caption{How to represent attributes}
    \label{tab:AttributeExamp}
\end{table}

One \textbf{pitfall} is to add attributes to your model that you are not concerned about instead of randomizing the values of that attribute at the concrete test level to validate that it does not cause a problem to surface. Consider the following exercise.

\begin{exercise}  \label{Exercise:AddedFunction}
Consider a function that calculates the quadratic equations $ax^2+b^x+c = 0$ locally and without an internal state.  Given a set of inputs $a, b$, and $c$ the function returns the solutions of the equation. In addition, the system can have remote communication with another backup system.  The remote communication could be either up or down. A tester added a remote communication attribute to the model that indicates whether or not the remote storage is accessible. Which of the following statements best represent the situation. 

\begin{enumerate}
\item The function is correct regardless of whether or not the backup storage system is accessible.  Thus, that attribute should be removed from the model and the state of the remote system can be randomly changed at the testing level if desired. 
\item The model needs to be comprehensive and include all possible attributes.  Therefore, it is good that the remote system accessibility attribute was added to the model.
\item We are concerned that we are making a mistake therefor we add the accessibility attribute to the model.
\item The accessibility attribute should be moved to the testing level and randomized on.
%DEB: how is this last option different thna the first option?
\end{enumerate}
See solution \ref{Exercise:AddedFunction:solution}
\end{exercise}

\section{Values}

Values are a collection of numbers, symbols, characters and strings that form a discrete unit which can be assigned to an attribute.  Each attribute has a set of possible values.  Questions that can be asked to help identify values for a given attribute follow.  

\begin{enumerate}
    \item What values are specified in the requirements or design?
    \item What values will alter the processing or results of the system?
    \item What values will drive different business rules or business logic?
    \item What values are not valid?
    \item What values typically cause problems?
\end{enumerate}

Let's look at an example to help clarify how answering these questions can drive interesting values for our attributes.

\begin{example}
We will continue using the example of our file system.  We need to test a file system implementation in which we are focusing on manipulation of files in the system.  We will look specifically at the attributes that make up a file. 

\begin{enumerate}
    \item Name of File
    \item Location of File
    \item Authority of File
    \item State of File
    \item Open File Authority
    \item Open File State
    \item Read File State
\end{enumerate}

Now that we have identified attributes, we start asking the questions listed above.  The first question is "What values are specified in the requirements or design"?  The following table \ref{tab:ValueExamp} flushes out the possible values that are interesting.

\begin{table}[H]
    \centering
    \begin{tabular}{|l|l|}
    \hline
    \cellcolor{lightgray}\textbf{Attribute}&\cellcolor{lightgray}\textbf{Values} \\
    \hline
    \multirow{4} {7em} {Name of File} &Maximum Length, minimum length, medium length \\
    &Mixed case, upper case, lower case \\
    &Special characters \\
    &Already exists, does not exist\\
    \hline  
         Location of File& Root directory, sub-directory, hidden directory\\
    \hline
         Authority of File& Authorized, unauthorized \\
    \hline
         State of File& Exists, does not exist \\
    \hline
         Open File Authority& Read only, write only, read and write\\
     \hline 
         Open File State& File exists, file does not exist \\
     \hline
        Read File State& File is opened, File is closed, File does not exist \\
     \hline
    \end{tabular}
    \caption{What values are specified in the requirements or design?}
    \label{tab:ValueExamp}
\end{table}

The next set of questions will help us to hi-lite values that are interesting.
\begin{itemize}
    \item "What values will alter the processing or results of the system?" 
    \item "What values will drive different business rules or business logic?" 
    \item "What values are not valid?" 
    \item "What values typically cause problems?" 
\end{itemize}

In the table below, the bold text highlights the values that are more interesting and are identified when answering the questions.  For example, we may have had problems with hidden directories. So, we identify hidden directories as an interesting value.  We are also concerned with security therefore authorized and unauthorized access is of interest.

\begin{table}[H]
    \centering
    \begin{tabular}{|l|l|}
    \hline
    \cellcolor{lightgray}\textbf{Attribute}&\cellcolor{lightgray}\textbf{Values} \\
    \hline
    \multirow{4} {7em} {Name of File} &Maximum Length, minimum length, medium length \\
    &Mixed case, upper case, lower case \\
    &Special characters \\
    &Already exists, does not exist\\
    \hline  
         Location of File& Root directory, sub-directory, \textbf{hidden directory}\\
    \hline
         Authority of File& \textbf{Authorized}, \textbf{unauthorized} \\
    \hline
         State of File& Exists, \textbf{does not exist} \\
    \hline
         Open File Authority& \textbf{Read only}, write only, read and write\\
     \hline 
         Open File State& File exists, \textbf{file does not exist} \\
     \hline
        Read File State& File is opened, File is closed, \textbf{File does not exist} \\
     \hline
    \end{tabular}
    \caption{How to Identify Interesting values}
    \label{tab:ValueExamp2}
\end{table}
\end{example}

Attributes may have prohibitive or infinite sets of values.  This makes it difficult to represent all possible values when modeling the testing space.  To solve this problem, we break down the potential values into representative classes.  The following table \ref{tab:RepresentValue} illustrates how this can be done.  For example, the size of a file name may be a range from 1 to 25 characters.  Instead of picking each value, we choose minimum, maximum and medium for possible values of the length of a filename.  This is generalized into the subdomain partition technique which we explain in the next subsection.

\begin{table}[H]
    \centering
    \begin{tabular}{|l|l|}
    \hline
    \cellcolor{lightgray}\textbf{Class}&\cellcolor{lightgray}\textbf{Example} \\
    \hline
    \multirow{4} {9em} {Numeric} &0, 10, single digit \\ 
    &upper bound, upper bound +1, lower bound, lower bound -1 \\
    &valid, not valid \\
    &minimum, maximum, middle \\
    &small, medium, large\\
    \hline  
         String &null, empty\\
    &starts with number, special character \\
    &maximum length, minimum length, maximum length +1 \\
    &valid value, not valid value, default value\\
    \hline
         Boolean &true, false, on, off, 0, 1 \\
    \hline
         Discrete & item number in an order system \\
    &element name to add to queue \\
    &valid, not valid \\
    \hline
         Ranges and boundaries &valid, not valid \\
    &does not exist, not enough authority \\
    &less than enough space, just enough space, more than enough space \\
     \hline 
    \end{tabular}
    \caption{How to Represent values}
    \label{tab:RepresentValue}
\end{table}

\subsection{Subdomain partition}

Subdomain partition is a way to reduce the number of values an attribute has based on concerns that are driven by boundaries.  An example will clarify the concept.  Assume you have an attribute, $x$, that is a number, and a test hint is available that the software under test makes a decision based on whether or not $x$ is bigger than 0.  Based on that test hint we are concerned that the software under test condition may be incorrect and decide to model $x$ values as $\{biggerThanZero, zero, lessThanZero\}$.  What we have done is implicitly partitioned the possible values of $x$ to $x < 0$, $x = 0$ and $x > 0$ hence the name subdomain partition. 

The subdomain, $x$ to $x < 0$, $x = 0$ and $x > 0$ in the example above are not overlapping.  In fact, this need not always be the case.  We may choose to have subdomains that overlap. When the abstract tests represented by the model are translated to concrete tests that execute against the software under test, the best practice is to randomly choose a value from the subdomain.  For example for the subdomain $lessThanZero$ we will choose a number that is less then zero each time we execute the test. 

Typically, the subdomain partition result in a set of possible values and we choose a representative.  For example, a "big file" subdomain refers to files in some range, say bigger than 1GB, and we randomly choose at concrete test generation time a file size that is in that range.  
A special case occurs when a specific value is of concern.  The maximum integer value and in such a case, test generation is reduced to picking exactly that value.

\subsection{How do concerns drive representation of attribute's value?}

There are many ways to model the values of a given attribute.  In fact, one may also separate an attribute to a few different attributes or keep them together.  For example, the filename attribute mentioned above may include in its value representation length, type of characters, mix case, and upper cases,  These values can be represented as part of the value name or they could be added as separate attributes.  For example, the length of the file name could be a separate attribute with values of maximum, minimum and medium.  The person creating the model may be confused as to when to separate the attributes or when to "put them together" in the value name of a single attribute. For instance the filename attribute may have a value of mixedcase-maxlen which is mixing two fundamental values that capture the characters of the file and how big the filename is.  In this section we explain the trade-offs that need to be taken into account when making such decisions.

Your modeling should be driven by concerns.  Assume a file can be be located in the root directory, sub-directory or hidden directory which is represented as another attribute in our model, and further assume that you are concerned about files with different file name length that resides in the different type of possible directories, you probably want to have the filename length attribute as a separate attribute.  This will ensure that the model will design tests that exercise the interaction of type of directories and type of filename length. 

The trade-off that we are facing is now becoming clear.  As a result of adding the filename length as a separate attribute, we have increased the number of combinations that the model represents.  This should only be done if we are concerned about these combinations.  If we are not concerned about the combinations of directory type and filename length there is no need to represent them as separate attributes and increase the complexity of the model by increasing the number of its combinations.  In fact, if we blindly represent everything as separate attributes we will needlessly make our testing effort more expensive and will probably not reveal more bugs.

\section{Constraints}

Constraints in a model represent impossible attribute/value combinations or they can also represent uninteresting attribute/value combinations.  Constraints are used to eliminate combinations of values that do not make logical sense.  Care needs to be taken when defining constraints because accidentally over-restricting a model will generate a test plan that is missing necessary values.

As with identifying attributes and values, there are a set of questions that can be asked to help define the constraints to be added to a model.  The questions are as follows.

\begin{itemize}
    \item "What combinations of attributes and values are not possible?"
    \item "What attribute values are only valid when other values are selected?"
    \item "What business rules prohibit certain attribute combinations?"
    \item "What parameter values can not interact?"
    \item "What states of the system are impossible to reach?"
    \item "What values are illegal and must be handled?"
\end{itemize}

Continuing with our previous file system example, the following table \ref{tab:PossibleConstraints} shows possible constraints for a set of attributes.
\begin{table}[H]
    \centering
    \begin{tabular}{|l|l|}
    \hline
    \cellcolor{lightgray}\textbf{Attribute}&\cellcolor{lightgray}\textbf{Possible constraint} \\
    \hline
    Name and Location of File &Max Length of name is not allowed when Location \\
    &of file is in a sub-directory \\ 
    \hline  
         Location of File and Authority of File &Unauthorized files are not allowed in the Root directory\\
    \hline
         Open File Authority and RW Action &Write to a file action is only allowed \\
         &when Open File Authority is Write Only or Read/Write \\
     \hline 
    \end{tabular}
    \caption{Example of Possible constraints}
    \label{tab:PossibleConstraints}
\end{table}

Once constraints have been identified, a question that may be asked is "how to represent constraints in the model?"  Most CTD tools allow for constraints to be represented with logic statements.  Using our file system example once again we can represent a constraint as follows.

\begin{example}
    ((OpenFileAuthority=WO) OR (OpenFileAuthority=RW)) AND (WriteAction=true)
\end{example}

Creating a model is an iterative process.  Once you have identified the attributes, values and constraints that are of interest, you then refine the attributes, clarify ambiguities and close gaps within the model definition.  When the model is complete, then the attributes, values and constraints can be entered into various tools that will do the test reduction.

\chapter{Coverage and Reduction} \label{ChapterCoverageReduction}
The number of possible tests of all combinations of attributes and values defined in the CTD model is called the full Cartesian product. The Full Cartesian product may produce a set of tests that is still too big to manage given resource and time to perform the test.  We add in constraints to limit the tests, yet that may still leave us with a large set of possible tests to execute.  When the testing space is so large, we risk running away from defining a model because the set of possible tests is too big.  Yet, if we do not model the testing space, we are left with ad-hock testing which runs the risk of missing interactions between attribute's values and thus missing identification of possible problems. 

If we can not cover all the tests produced from a defined model, then reduction of the tests is needed.  Evidence indicates that the majority of problems in a software system are triggered by only a small set of interactions.  Based on that, we look at n-way coverage.  Where n is the number of attribute and value combinations that we want the tests to cover.  Evidence shows that 2 or 3 interactions is enough.  

The following example shows a pairwise (or 2-way) test on an Application Programming Interface (API) that has the following input parameter attributes and possible values.
\begin{example}
Table \ref{tab:APIparmValue} is a description of the model of a sample API that has 7 parameters and each parameter has 2 possible values.
    \begin{table}[H]
    \centering
    \begin{tabular}{|l|l|}
    \hline
    \cellcolor{lightgray}\textbf{Parameter}&\cellcolor{lightgray}\textbf{Possible values} \\
    \hline
         Name& Bob, Fred\\
    \hline  
        Shape& Star, Oval\\
    \hline
         Color& Red, Blue \\
    \hline
         Integer& 1, 2 \\
    \hline
         Animal& Dog, Cat \\
     \hline 
         Age& 2, 7 \\
     \hline
        Location& USA, UK \\
     \hline
        Gender& M, F \\
     \hline
    \end{tabular}
    \caption{API parameters and values}
    \label{tab:APIparmValue}
\end{table}

If we were to look at all the combinations given the above model, we would have $2^8$ or 256 tests as part of the full Cartesian product.

If 2-way reduction was applied to this model, it could be reduced from 256 down to 7 test programs.
\begin{table}[H]
    \centering
    \begin{tabular}{|l|l|l|l|l|l|l|l|l|}
    \hline
    \cellcolor{lightgray}\textbf{Test} &\cellcolor{lightgray}\textbf{Name} &\cellcolor{lightgray}\textbf{Shape}&\cellcolor{lightgray}\textbf{Color}&\cellcolor{lightgray}\textbf{Integer}&\cellcolor{lightgray}\textbf{Animal}&\cellcolor{lightgray}\textbf{Age}&\cellcolor{lightgray}\textbf{Location}&\cellcolor{lightgray}\textbf{Gender} \\
    \hline
         1& Bob& Star& Blue& 1& Cat& 2& UK& F\\
    \hline  
        2& Fred& Oval& Blue& 1& Dog& 7& UK& M\\
    \hline
        3& Bob& Oval& Red& 2& Cat& 2& USA& F \\
    \hline
        4& Bob& Star& Red& 1& Dog& 2& USA& F\\
    \hline
        5& Fred& Star& Blue& 2& Cat& 2& USA& M \\
     \hline 
        6& Bob& Star& Red& 2& Dog& 7& UK& M \\
     \hline
        7& Fred& Oval& Red& 1& Cat& 7& USA& F \\
     \hline
    \end{tabular}
    \caption{2-way test set}
    \label{tab:2wayTestSet}
\end{table}

If we were to look at the combinations of 2 attributes, Name and Shape, we would see as depicted in table \ref{tab:NameShape} that we have all the values of Shape with the Name value of Bob, highlighted in yellow with tests 1 and 3 and all the values of Shape with the Name value of Fred, highlighted in orange with tests 5 and 7.
\begin{table}[H]
    \centering
    \begin{tabular}{|l|l|l|l|l|l|l|l|l|}
    \hline
    \cellcolor{lightgray}\textbf{Test} &\cellcolor{lightgray}\textbf{Name} &\cellcolor{lightgray}\textbf{Shape}&\cellcolor{lightgray}\textbf{Color}&\cellcolor{lightgray}\textbf{Integer}&\cellcolor{lightgray}\textbf{Animal}&\cellcolor{lightgray}\textbf{Age}&\cellcolor{lightgray}\textbf{Location}&\cellcolor{lightgray}\textbf{Gender} \\
    \hline
         1& \cellcolor{yellow}Bob& \cellcolor{yellow}Star& Blue& 1& Cat& 2& UK& F\\
    \hline  
        2& Fred& Oval& Blue& 1& Dog& 7& UK& M\\
    \hline
        3& \cellcolor{yellow}Bob& \cellcolor{yellow}Oval& Red& 2& Cat& 2& USA& F \\
    \hline
        4& Bob& Star& Red& 1& Dog& 2& USA& F\\
    \hline
        5& \cellcolor{orange}Fred& \cellcolor{orange}Star& Blue& 2& Cat& 2& USA& M \\
     \hline 
        6& Bob& Star& Red& 2& Dog& 7& UK& M \\
     \hline
        7& \cellcolor{orange}Fred& \cellcolor{orange}Oval& Red& 1& Cat& 7& USA& F \\
     \hline
    \end{tabular}
    \caption{Combinations of Name and Shape attributes}
    \label{tab:NameShape}
\end{table}

If we were to look at the combinations of 2 attributes, Name and Color, we would see as depicted in table \ref{tab:NameColor} that we have all the values of Color with the Name value of Bob, highlighted in yellow with tests 1 and 6 and all the values of Color with the Name value of Fred, highlighted in orange with tests 5 and 7.
\begin{table}[H]
    \centering
    \begin{tabular}{|l|l|l|l|l|l|l|l|l|}
    \hline
    \cellcolor{lightgray}\textbf{Test} &\cellcolor{lightgray}\textbf{Name} &\cellcolor{lightgray}\textbf{Shape}&\cellcolor{lightgray}\textbf{Color}&\cellcolor{lightgray}\textbf{Integer}&\cellcolor{lightgray}\textbf{Animal}&\cellcolor{lightgray}\textbf{Age}&\cellcolor{lightgray}\textbf{Location}&\cellcolor{lightgray}\textbf{Gender} \\
    \hline
         1& \cellcolor{yellow}Bob& Star& \cellcolor{yellow}Blue& 1& Cat& 2& UK& F\\
    \hline  
        2& Fred& Oval& Blue& 1& Dog& 7& UK& M\\
    \hline
        3& Bob& Oval& Red& 2& Cat& 2& USA& F \\
    \hline
        4& Bob& Star& Red& 1& Dog& 2& USA& F\\
    \hline
        5& \cellcolor{orange}Fred& Star& \cellcolor{orange}Blue& 2& Cat& 2& USA& M \\
     \hline 
        6& \cellcolor{yellow}Bob& Star& \cellcolor{yellow}Red& 2& Dog& 7& UK& M \\
     \hline
        7& \cellcolor{orange}Fred& Oval& \cellcolor{orange}Red& 1& Cat& 7& USA& F \\
     \hline
    \end{tabular}
    \caption{Combinations of Name and Color attributes}
    \label{tab:NameColor}
\end{table}

To belabor this point, this last example looks at the combinations of 2 attributes, Integer and Animal.  We see depicted in table \ref{tab:IntegerAnimal} that we have all the values of Animal with the Integer value of 1, highlighted in yellow with tests 4 and 7 and all the values of Animal with the Integer value of 2, highlighted in orange with tests 3 and 6.
\begin{table}[H]
    \centering
    \begin{tabular}{|l|l|l|l|l|l|l|l|l|}
    \hline
    \cellcolor{lightgray}\textbf{Test} &\cellcolor{lightgray}\textbf{Name} &\cellcolor{lightgray}\textbf{Shape}&\cellcolor{lightgray}\textbf{Color}&\cellcolor{lightgray}\textbf{Integer}&\cellcolor{lightgray}\textbf{Animal}&\cellcolor{lightgray}\textbf{Age}&\cellcolor{lightgray}\textbf{Location}&\cellcolor{lightgray}\textbf{Gender} \\
    \hline
         1& Bob& Star& Blue& 1& Cat& 2& UK& F\\
    \hline  
        2& Fred& Oval& Blue& 1& Dog& 7& UK& M\\
    \hline
        3& Bob& Oval& Red& \cellcolor{orange}2& \cellcolor{orange}Cat& 2& USA& F \\
    \hline
        4& Bob& Star& Red& \cellcolor{yellow}1& \cellcolor{yellow}Dog& 2& USA& F\\
    \hline
        5& Fred& Star& Blue& 2& Cat& 2& USA& M \\
     \hline 
        6& Bob& Star& Red& \cellcolor{orange}2& \cellcolor{orange}Dog& 7& UK& M \\
     \hline
        7& Fred& Oval& Red& \cellcolor{yellow}1& \cellcolor{yellow}Cat& 7& USA& F \\
     \hline
    \end{tabular}
    \caption{Combinations of Integer and Animal attributes}
    \label{tab:IntegerAnimal}
\end{table}

\end{example}

At this point, it should be clear how the full Cartesian product of tests depicted by a model comprising of attributes and values can be optimized by doing n-wise reduction.  Next we will look at how identifying constraints in conjunction with the n-wise reduction can further optimize the number of tests.

\chapter{How to define constraints and coverage to best direct your tests}

There are two approaches regarding \textbf{why} you would add constraints to a model.  The first is to add constraints due to invalid or uninteresting attribute/value combinations.  The second reason for adding constraints is to additionally minimize the testing space after the test reduction has occurred. It is good practice to document the reason for adding the constraint otherwise the model becomes unreadable. Let us explore the second reason.
\begin{exercise}  \label{Exercise:Constraint1}
    Suppose we have a model defined below. 
\begin{table}[H]
    \centering
    \begin{tabular}{|l|l|}
    \hline
    \cellcolor{lightgray}\textbf{Attribute}&\cellcolor{lightgray}\textbf{Values} \\
    \hline
         attribute1 & value1, value2, value3, value4, value5, value6, value7, value8, value9 \\
    \hline  
         attribute2 & value1, value2, value3, value4, value5, value6, value7\\
    \hline
        attribute3 & value1, value2, value3, value4, value5\\
    \hline
        attribute4 & value1, value2\\
     \hline 
    \end{tabular}
    \caption{Model 1}
    \label{tab:ExerciseModelReduction}
\end{table}
    When this model is run through CTD pair-wise reduction we get the following 63 tests
%\begin{table}[H]
\begin{longtable}{|l|l|l|l|l|}
    %\centering
    %\begin{tabular}{|l|l|l|l|l|}
    \hline
    \cellcolor{lightgray}\textbf{Test}&\cellcolor{lightgray}\textbf{attribute1}&\cellcolor{lightgray}\textbf{attribute2}&\cellcolor{lightgray}\textbf{attribute3}&\cellcolor{lightgray}\textbf{attribute4}\\
    %\hline
    %Test& attribute1& attribute2& attribute3& attribute4\\
    \hline
1& value5 & value4 & value5 & value1 \\
\hline
2& value4 & value6 & value3 & value1 \\
\hline
3& value3 & value6 & value5 & value2 \\
\hline
4& value3 & value2 & value4 & value1 \\
\hline
5& value5 & value3 & value4 & value2 \\
\hline
6& value8 & value5 & value1 & value1 \\
\hline
7& value1 & value3 & value2 & value1 \\
\hline
8& value9 & value1 & value1 & value2 \\
\hline
9& value4 & value5 & value2 & value2 \\
\hline
10& value1 & value4 & value3 & value2 \\
\hline
11& value9 & value7 & value3 & value1 \\
\hline
12& value2 & value1 & value5 & value1 \\
\hline
13& value7 & value7 & value2 & value2 \\
\hline
14& value2 & value2 & value2 & value2 \\
\hline
15& value6 & value2 & value1 & value2 \\
\hline
16& value8 & value4 & value2 & value2 \\
\hline
17& value6 & value5 & value3 & value1 \\
\hline
18& value7 & value1 & value4 & value1 \\
\hline
19& value7 & value3 & value3 & value1 \\
\hline
20& value5 & value7 & value1 & value2 \\
\hline
21& value9 & value6 & value2 & value1 \\
\hline
22& value2 & value5 & value4 & value2 \\
\hline
23& value3 & value3 & value1 & value2 \\
\hline
24& value8 & value7 & value4 & value2 \\
\hline
25& value5 & value2 & value3 & value1 \\
\hline
26& value3 & value1 & value3 & value1 \\
\hline
27& value2 & value6 & value1 & value1 \\
\hline
28& value6 & value4 & value4 & value1 \\
\hline
29& value9 & value3 & value5 & value1 \\
\hline
30& value1 & value6 & value4 & value1 \\
\hline
31& value8 & value2 & value5 & value1 \\
\hline
32& value1 & value7 & value5 & value1 \\
\hline
33& value7 & value5 & value5 & value2 \\
\hline
34& value7 & value4 & value1 & value1 \\
\hline
35& value6 & value1 & value2 & value1 \\
\hline
36& value9 & value2 & value4 & value1 \\
\hline
37& value5 & value6 & value2 & value2 \\
\hline
38& value8 & value3 & value3 & value1 \\
\hline
39& value1 & value2 & value1 & value1 \\
\hline
40& Value4 & value3 & value1 & value2 \\
\hline
41& Value4 & value4 & value4 & value1 \\
\hline
42& Value2 & value7 & value3 & value1 \\
\hline
43& Value4 & value7 & value5 & value2 \\
\hline
44& Value3 & value5 & value2 & value2 \\
\hline
45& Value6 & value6 & value5 & value1 \\
\hline
46& Value4 & value1 & value4 & value1 \\
\hline
47& Value8 & value6 & value4 & value2 \\
\hline
48& Value4 & value2 & value1 & value1 \\
\hline
49& Value2 & value4 & value5 & value2 \\
\hline
50& Value7 & value2 & value4 & value1 \\
\hline
51& Value1 & value5 & value4 & value1 \\
\hline
52& Value3 & value4 & value3 & value1 \\
\hline
53& Value7 & value6 & value4 & value2 \\
\hline
54& Value8 & value1 & value1 & value2 \\
\hline
55& Value3 & value7 & value3 & value1 \\
\hline
56& Value5 & value5 & value5 & value1 \\
\hline
57& Value6 & value3 & value4 & value1 \\
\hline
58& Value1 & value1 & value1 & value2 \\
\hline
59& Value2 & value3 & value5 & value2 \\
\hline
60& Value6 & value7 & value5 & value2 \\
\hline
61& Value5 & value1 & value2 & value1 \\
\hline
62& Value9 & value4 & value2 & value1 \\
\hline
63& Value9 & value5 & value2 & value1 \\
\hline
    %\end{tabular}
    \caption{Model Reduction Tests}
    \label{tab:ExerciseModelReductionTests}
\end{longtable}
    Next supposed that each test will take 2 minutes to execute and you only have 60 minutes before the weekend starts.  You would like to reduce the testing space even further to accommodate this time constraint. What is the least number of constraints to the CTD model to get the largest test reduction? 
    \begin{enumerate}  
    \item Remove them all, who needs to test anything one hour before the weekend?
    \item Randomly pick 30 tests.
    \item Remove values from attribute4, then attribute3, then attribute2, then attribute1 until we are left with 30 tests
    \item Remove values from attribute1, then attribute2, then attribute3 then attribute4 until we are left with 30 tests
    \end{enumerate}
% The above question: It is more efficeint to remove values from the attributes that have the most values. So, removing from attribute1 then attribute2 would get you to 30 in the least amount of steps 
%TBC: we can take the model and generate all the constraints and put these as examples in the back of the book
%TBC: we can be more specific and say remove value6-value9 from attribute1 and value 6-value7 from attribute2Constraint
See solution \ref{Exercise:Constraint1:solution}
\end{exercise}

\chapter{A deeper dive into CTD - principles and algorithms}
%Slide 49
Next we consider how to implement a CTD tool.  We will focus on a specific possible implementation using a data structure called Binary Decision Diagram (BDD).  This is not the only option but it will serve the purpose to make the discussion concrete.  The first part of the discussion motivates the choice of the data structure.  It outlines requirements that would apply to any design of a CTD tool.

In the process of designing a CTD model one needs to capture, in addition to the attributes and values, the legal and illegal combinations of attributes and values.  This reflects either the semantic of the system or testing interests and concerns. Thus, a good CTD tool will need to be able to explicitly represent illegal and legal combinations and easily enable their exploration and the removal and addition of restrictions to facilitate the model development process. 

The requirement of explicit representation of legal and illegal combinations does not only apply to the entire set of attributes.  In the process of designing a CTD model one needs to view projections of the entire Cartesian product represented by the model.  In other words, one may be interested in viewing a subset of the attributes on all of the legal combinations of values that apply to that subset.  To sum up the tool is required to enumerate subsets of the model Cartesian product the model represents for various purposes that facilitate the modeling process.

Another requirement of a CTD tool is to be able to provide an optimal subset of the Cartesian product that meets some given interaction coverage requirement. To setup expectations it is not always feasible to obtain the smallest possible subset as there are some theoretical limitations to that task but meaningful reduction, in practice, by an order of magnitude should be achieved by the tool.  For example, models that represents on the order of hundreds of thousands of combinations should be reduced by the tool to a set of dozen of combinations that meets, say, 2-way coverage.   

%SLIDE 50

To setup the expectations for a CTD tool one needs to understand in general finding a minimal subset of the legal combinations of a model is hitting the theoretical boundaries of computations, in other words it is considered a hard problem.  Instead, we define a greedy heuristic that provides a small plan that meets the requirements.   The plan is not necessarily the optimal possible plan but in practice it enables most industrial applications of the technique and provides a sizeable reduction in the test effort.  

Now that we setup our expectations of the tool and outlines its most crucial requirements, namely to handle many types of subsets of the Cartesian product of test parameters and their interaction gracefully we are nearly ready to outline a possible approach to the design of a CTD tool.  

To recap we made a list of the categories that the tool will need to handle and clarify why each category boils down to the need to represent a subset of the test Cartesian product defined by some condition. 
\begin{itemize}
\item A legal test.  A legal test is a combination of test parameters that can be executed against the system. What is a legal test is defined in terms of conditions on the interactions between parameters.  For example, once a generation parameter named greedy is used another generation parameter named temperature is not applicable.  This defines types of generation that can be used and exclude others.  Each generation type translates to a legal test. 
\item Illegal tests.  Illegal test is a test that is not legal. Once it is clear that the legal tests are a subset of the Cartesian product of test parameters defined by a condition, it is also clear that the illegal tests are a subset of the Cartesian product of test parameters defined by the negation of the condition that defines the legal tests. 
\item Projections.  Projection here is defined as the subset of the Cartesian product for which certain test parameters assume some specific value.   For example, a subset of a CTD model that represent testing the $open()$ of a file with the file type being set to a regular file and the user permission is set to a read permission.  Clearly by its definition projection is a subset defined by a condition.  
\item Interaction coverage requirements.  Interaction coverage requirement requires that a set of the test attributes will assume a given value.  For example, in the case of $2$-way coverage and for the previous example, one of the resulting requirements is that the file being opened is regular and that the permission that the user have to the file is a read permission.  Such an interaction coverage  requirement can be viewed as a projection.  Thus, interaction coverage requirements can be defined using conditions.
\item Set of legal tests.  A set of legal tests is obtained by taking together legal tests.  As discussed above each legal test is determined by some condition, hence a set of legal tests is determined by the taking either one of the conditions that determine the individual legal tests.   To give a concrete example, if one legal test is defined as $t_1 = (openType = read, permission = rw, fileType = regular)$ and another legal test is defined by the condition $t_2 = (openType = write, permission = w, fileType = socket)$ then the set of legal tests may include $t_1$  and $t_2$ and be defined by the condition $t_1 \lor t_2$.  
\item A set of existing tests is just a set of tests that were already executed against the system under test and are thus defined by conditions as discussed in the previous bullet. 
\end{itemize}

The tool should also be able to easily support operations on Boolean functions.  As the design of the model occurs, the restrictions are continually changing and being updated.   In addition, coverage requirements are modified.  This ongoing modification of constraints and coverage requirements, as well as the addition and removal of parameters and their associated values are continually occurring even after the model has been successfully used in testing the system under test.  This is the case as a software components are continually enhanced with new features and need to be continually tested and regressed.  As a result the Boolean functions that defined the constraints and coverage requirements are continually changing.  To smoothly support that process the tool needs to support easy implementation of logical operations such as conjunction so that the constraints and coverage requirements are put together to represent the desired space that is to be tested.  

Another source of continued need to apply logical operations to different logical expression arises from the need to represent projections of the space of possible tests.  A projection is obtained by specifying the value of a subset of the attributes.  Again doing that requires "putting together" the required projection logical conditions with the logical constraints that define the legal part of the testing space.  As a result the requirement for exploration of the testing space through projection leads to the need to support different logical operations on Boolean functions.  

\begin{example}
Consider the following example. We are given a 
test space that consists of a Cartesian product of $x1,x2, x3, x4$ of type Boolean.  The legal combinations are combinations with at
least one $1$ which is represented by the Boolean
function A, e.g., $1011$, $1001, \ldots$.
We would like to obtain a legal test that covers the
interaction $x1 = 0$ and $x2 = 0$ which is
represented by the Boolean function B.  The tool needs to calculate the answer to that requirement that is $0010, 0001, 0011$.  The answer is obtained by taking the conjunction of the two Boolean functions $A \land B$.   
To answer the required query the tool needs to calculate $A \land B$ and then enumerate all the legal tests that meet that requirement. 

\end{example}

To sum up, in order to support an implementation of the CTD approach one needs a data structure that efficiently handles logical formulas and their associated operations such as conjunction, negation and enumeration of the combinations that evaluate to true for a given formula.   Binary Decision Diagrams (BDD) to the rescue!  BDDs are exactly such data structures as they efficiently represent logical formulas and execute their associated logical operations.  Although other implementations for the CTD approach are possible we will concentrate on BDDs in this chapter as we want to emphasise the need for such support of the design process of the CTD model. 

To implement the CTD approach the algorithm should obtain a set of combinations that meet the interaction coverage requirements. Ideally the set should be as small as possible.  The following two points are emphasised.  On the one hand it is not always possible to obtain a minimal set of combinations of parameters' values that meet the desired interaction coverage.  In addition, the support of the CTD modeling design process as described above is as important as getting a minimal set.  Assuming we have a CTD model with $10^9$ legal combinations.  This size of a model are in fact typical in practice.  If two different algorithms get us a set of parameters value combinations that meets the required interaction coverage of size on the order of $100$ it is probably not important which of the algorithms are used.  In practice it may be the case that either reduction would be sufficient to execute the entire set of parameters value combinations against the system under test.  In such a case we would prefer the algorithm that enables the design process of the CTD model.  BDD data structure does exactly that.

Boolean function are functions from $X_1\times, \ldots, \times X_n$ to $X$. 
$X_i$ and $X$ get values in $\{0, 1\}$ representing true and false respectively.  
A BDD is a data structure that represents a Boolean function.  More specifically a BDD is a Directed Acyclic Graph (DAG).  The nodes of the graph are associated with the Boolean function independent variables.  Each node, $x_i$, has two outgoing edges, namely, the $0$ labeled edge that indicates that the variable $x_i$ has the value $0$ and the $1$ labeled edge indicating the the variable $x_i$ has the value $1$.  The BDD has the following semantic.  In order to evaluate the Boolean function one traverses the BDD from the nodes that have no parent nodes to the leaf of the BDD.  The order of traversing the BDD matches some predefined order in which the Boolean function variables values are inspected.  It is customary to denote the order of variables inspection using the $<$ sign, e.g., $x_1 < x_2 <, \ldots,  < x_n$.  Whenever a node is inspected that represents independent variable $x_i$, if the value of $x_i$ is $1$ then the $1$ edge is followed, otherwise the $0$ edge is followed.  This process is continued until a leaf of the BDD is reached.  The leafs are associated with a $0$ or a $1$ which stands for the value of the Boolean function evaluation.  To clarify the above description, it is useful to study an evaluation example at this point. 

\begin{figure}
\begin{tikzpicture}[node distance=2cm]
\node (startBDD) [function] {X1};
\node (X2_1) [function, below of=startBDD, yshift=-2.0cm] {X2};
\node (X2_0) [function, right of=X2_1, xshift=2cm] {X2};
\node (X3_1) [function, below of=X2_0, xshift=5cm, yshift=-2.0cm] {X3};
\node (X3_0) [function, below of=X2_1, xshift=5cm, yshift=-1.0cm] {X3};
\node (End0) [process, below of=X3_0, xshift=-5cm, yshift=-1.5cm] {0};
\node (End1) [process, below of=X3_1, xshift=-5cm, yshift=-0.5cm] {1};

\draw [arrow] (startBDD) -- node[anchor=east] {1} (X2_1);
\draw [dasharrow] (startBDD) -- node[anchor=east] {0} (X2_0);
\draw [dasharrow] (X2_1) -- node[anchor=east] {0} (End0);
\draw [arrow] (X2_1) -- node[anchor=east] {1} (End1);
\draw [dasharrow] (X2_0) -- node[anchor=east] {0} (X3_0);
\draw [arrow] (X2_0) -- node[anchor=east] {1} (X3_1);
\draw [arrow] (X3_0) -- node[anchor=east] {1} (End0);
\draw [dasharrow] (X3_0) -- node[anchor=east] {0} (End1);
\draw [arrow] (X3_1) -- node[anchor=west] {1} (End1);
\draw [dasharrow] (X3_1) -- node[anchor=west] {0} (End0);

\node (BDDtable) [textbox, right of=startBDD, xshift=9cm] {
\begin{tabular}{ c c c | c  } 
  x1 & x2 & x3 & f \\ 
  \hline
  0 & 0 & 0 & 1 \\ 
  0 & 0 & 1 & 0 \\ 
  0 & 1 & 0 & 0 \\ 
  0 & 1 & 1 & 1 \\ 
  1 & 0 & 0 & 0 \\ 
  1 & 0 & 1 & 0 \\ 
  1 & 1 & 0 & 1 \\ 
  1 & 1 & 1 & 1 \\ 
  
\end{tabular}
};

\end{tikzpicture}
\caption{\label{BDD}Binary Decision Diagram (BDD)} \label{BDD_1}
\end{figure}
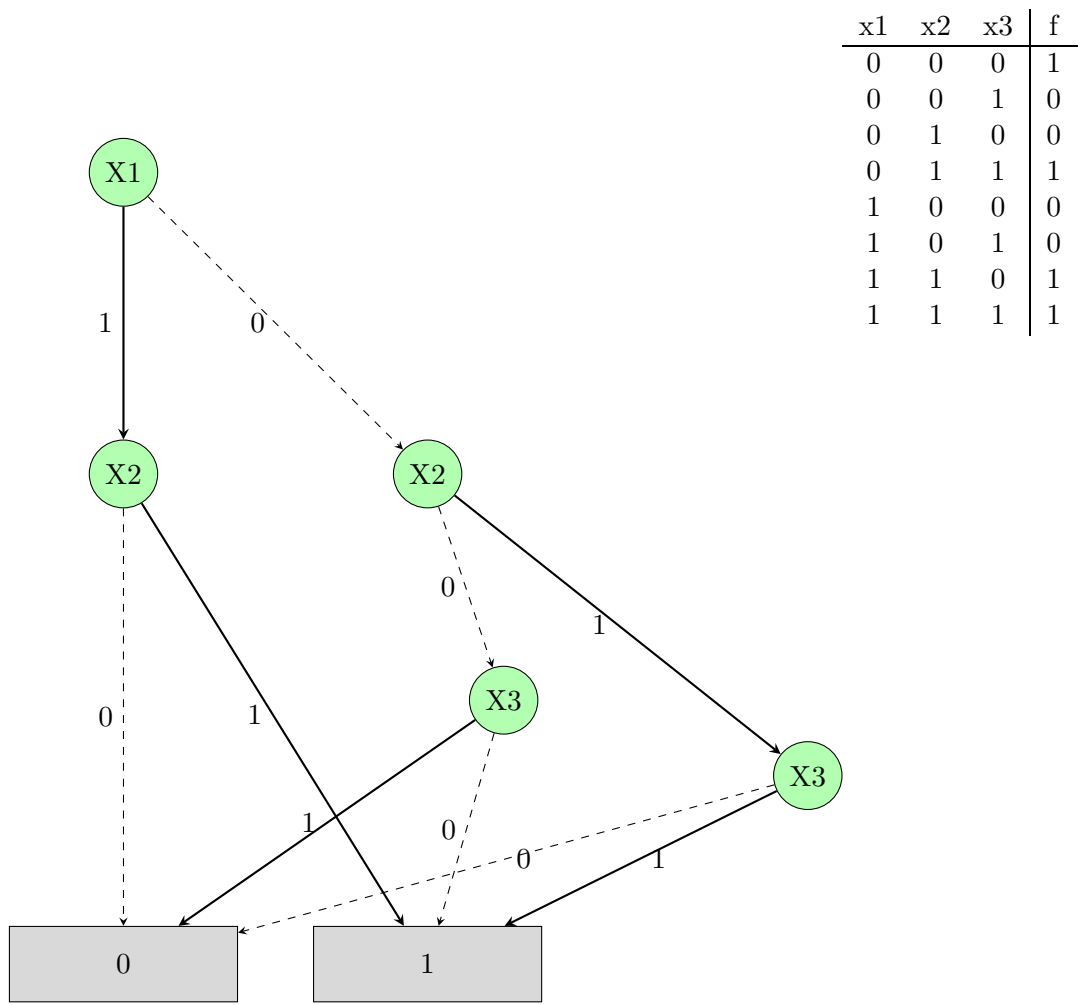

\begin{example}
A BDD is given in \ref{BDD}.  Assume that $x_1 = 0$, $x_2 = 1$, and $x_3 = 1$.  The evaluation starts at $x_1$ and then takes the $0$ edge reaching $x_2$. Next the $1$ edge is taken reaching $x_3$ and then the $1$ edge is taken again reaching the leaf of the BDD.  At that point we determine the the value of the Boolean function is $1$. 

If the outcome of the function is determined by the prefix of the variables already inspected, we 
“short cut” to the final answer.  See example in figure \ref{BDD}, 
if $x1=1$, and $x2=1$ the function outcome is 1
regardless of the value of $x3$.  We thus "short cut" to the leaf of the tree indicating that the function
value is $1$.

\end{example}

The BDD DAG is a recursive structure.  To see that, denote by $f(x1, \ldots , xn)$ the Boolean function.  Note that the sub DAG reached by a 1 arrow from x1 is a BDD of the Boolean function $f(x1, \ldots,xn)$ in which $x1$ is known to be $1$.  We denote that function as $f(x1=1, x2,\ldots,xn)$ or $f(x1=1)$ for short. 
Similarly, the sub DAG reached by the $0$ arrow from $x1$ is $f(x1=0)$. 
The Shannon expansion highlights this relation algebraically by stating that
$f(x1, \ldots ,xn) = x1f(x1=1)+(not~x1)f(x1=0)$.  If $x1 = 1$ then $(not~x1)=0$ hence
 $(not~x1)f(x1=0) = 0$.  In addition,
$x1f(x1=1) = 1f(x1=1) = f(x1=1)$. Hence, 
$x1f(x1=1)+(not~x1)f(x1=0) = 
f(x1=1)+0 =  f(x1=1)$.  A similar proof applies for the case of $x1 = 0$.  We sometimes refer to 
$x1f(x1=1)+(not~x1)f(x1=0)$ as
$ite(x1, f(x1 = 1), f(x1 = 0))$\footnote{ite stands for “if then else”.}

\begin{exercise}
Prove the Shannon expansion for the case of $x1 = 0$.
\end{exercise}

\begin{figure}

\begin{tikzpicture}[node distance=2cm]
\node (header_g) [textbox] {\underline{$g=x1 \wedge x2$}};
\node (start_g) [function, below of=header_g, yshift=-1.0cm] {X1};
\node (end_g_x1_0) [process, below of=start_g, yshift=-1.0cm] {0};
\node (g_x2_1) [function, right of=end_g_x1_0, xshift=2cm] {X2};
\node (End_g_x2_0) [process, below of=g_x2_1, xshift=-3cm, yshift=-1.0cm] {0};
\node (End_g_x2_1) [process, below of=g_x2_1, xshift=1cm, yshift=-1.0cm] {1};

\node (header_h) [textbox, right of=header_g, xshift=9cm] {\underline{$h=x1 \vee x2$}};
\node (start_h) [function, below of=header_h, yshift=-1.0cm] {x1};
\node (end_h_x1_1) [process, below of=start_h, xshift=3cm] {1};
\node (h_x2_0) [function, left of=end_h_x1_1, xshift=-2cm] {x2};
\node (end_h_x2_0) [process, below of=h_x2_0, xshift=-1cm, yshift=-0.7cm] {0};
\node (end_h_x2_1) [process, right of=end_h_x2_0, xshift=2cm] {1};

\draw [arrow] (start_g) -- node[anchor=east] {0} (end_g_x1_0);
\draw [dasharrow] (start_g) -- node[anchor=east] {1} (g_x2_1);
\draw [dasharrow] (g_x2_1) -- node[anchor=east] {1} (End_g_x2_1);
\draw [arrow] (g_x2_1) -- node[anchor=east] {0} (End_g_x2_0);

\draw [dasharrow] (start_h) -- node[anchor=east] {1} (end_h_x1_1);
\draw [arrow] (start_h) -- node[anchor=east] {0} (h_x2_0);
\draw [arrow] (h_x2_0) -- node[anchor=east] {0} (end_h_x2_0);
\draw [dasharrow] (h_x2_0) -- node[anchor=east] {1} (end_h_x2_1);

\end{tikzpicture}
\caption{BDDs for functions g and h} \label{BDD_function_g_h}
\end{figure}
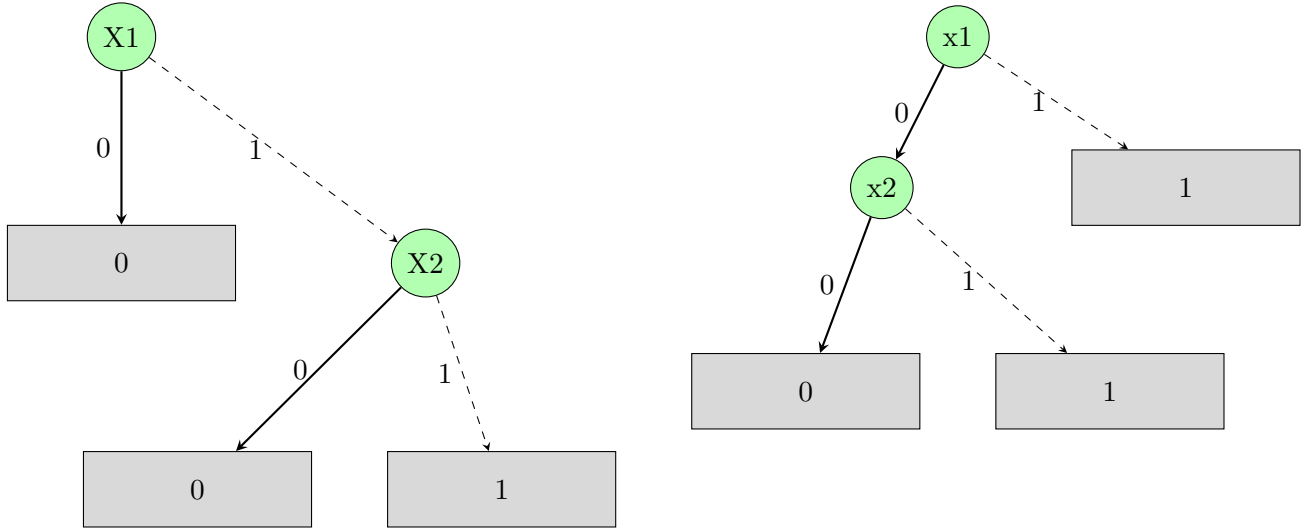

Next we demonstrate by example how we obtain the BDD of the and of to logical functions using the Shannon expansion.  We are given the functions $g$ and $h$ defined in Figure \ref{BDD_function_g_h}, and we proceed to calculate the BDD of $f = g \wedge h$  that is $f(x1,x2) = (x1 \wedge x2) \wedge (x1 \vee x2)$.  Using the Shannon expansion we have the following.  

\begin{figure}
\begin{tikzpicture}[node distance=2cm]
\node (header_f) [textbox] {\underline{$f(x1,x2) = (x1 \wedge x2) \wedge (x1 \vee x2)$}};
\node (start_f) [function, below of=header_f, yshift=-1.0cm] {X1};
\node (end_f_x1_0) [process, below of=start_f, yshift=-1.0cm] {0};
\node (f_x2_1) [function, right of=end_f_x1_0, xshift=2cm] {X2};
\node (End_f_x2_0) [process, below of=f_x2_1, xshift=-3cm, yshift=-1.0cm] {0};
\node (End_f_x2_1) [process, below of=f_x2_1, xshift=1cm, yshift=-1.0cm] {1};

\draw [arrow] (start_f) -- node[anchor=east] {0} (end_f_x1_0);
\draw [dasharrow] (start_f) -- node[anchor=east] {1} (f_x2_1);
\draw [dasharrow] (f_x2_1) -- node[anchor=east] {1} (End_f_x2_1);
\draw [arrow] (f_x2_1) -- node[anchor=east] {0} (End_f_x2_0);

\end{tikzpicture}
\caption{BDD for function f} \label{BDD_function_f}
\end{figure}
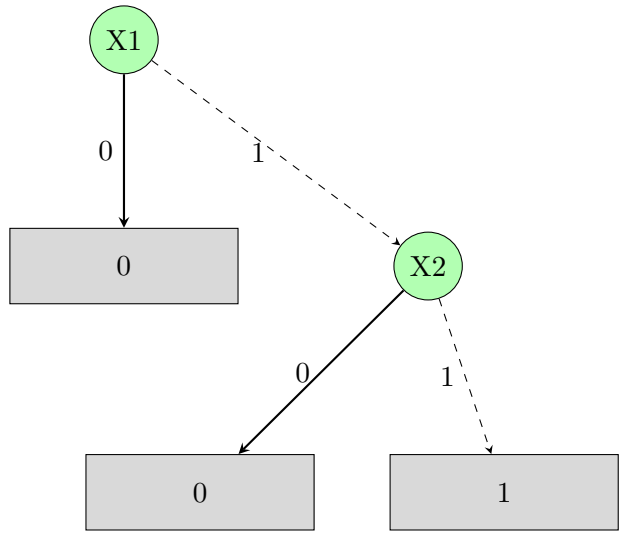

\begin{itemize}
\item 
$f(x1, x2) = ite(x1, f(x1 = 1), f(x1 = 0)) = $  
\item
$ite(x1, g(x1=1) \wedge h(x1 = 1), g(x1 = 0) \wedge h(x1 = 0)) = $
\item
$ite(x1, x2 \wedge 1, 0 \wedge x2) = ite(x1, x2 , 0)$  
\end{itemize}

which is the BDD depicted in figure \ref{BDD_function_f}.

\newpage

To summarize BDDs can support the implementation of combinatorial test design tools.  Specifically, they have the following desired characteristics.

\begin{itemize}
    \item Canonicity: given a Boolean function, and an order for the variables, there is only one BDD that represents
this function.  Due to this canonic representation, comparison of functions represented
by BDDs may be performed performed quickly in constant time.
    \item  Boolean operations such as conjunction, disjunction and negation can
also be computed quickly and efficiently.  Negation can be computed in constant time, conjunction
and disjunction in the worst case in time proportional to the multiplication of
the two sizes of the input BDDs.
   \item Counting the number of satisfying assignments and iterating over them
is another important examples of and efficient operations as well as obtaining a satisfying assignment from a BDD is efficient.
\end{itemize}

For more details on the implementation of combinatorial testing tools using BDDs see \cite{DBLP:conf/issta/SegallTF11}.  In addition, the interested reader may want to \href{http://www.lsv.fr/~schwoon/enseignement/verification/ws0910/bdd.pdf}{ review } that deep dive on BDDs.

%▪ Test optimization
%– Finding an optimal set of legal tests for a given
%required interaction coverage
%• While utilizing existing tests
%– Analyzing the interaction coverage of existing
%tests
%• Remove redundancy in existing tests given
%some required interaction coverage

%TBC - statistical assumption - bugs are small, the combinatorial optimization problem, solutions, the difference between free and not free models (with and without constraints), the advantage of BDD.  Talk about coverage requirement (interaction levels), examples of covering array

\chapter{Application of CTD to technical reviews}
Creating a CTD model to be used to assist in a code review is slightly different than creating a model to be used to generate test programs.  The purpose of these models are to generate thought and discussion in a review, not to create a perfect representation of the system.  In addition, such models typically captures interaction and states that may have been overlooked, thus finding issues during the review process.  As a side effect of the review, such CTD review models can also be further developed and be made more complete so that they be used to derive tests.  

\begin{figure}
\begin{tikzpicture}[node distance=2cm]
\node (client) [textbox] {\textbf{\underline{client}}};
\node (start) [startstop, below of=client] {Pick first element from queue};
\node (dec1) [decision, below of=start, yshift=-0.5cm] {end of queue?};
\node (stop) [startstop, below of=dec1, yshift=-0.5cm] {end of processing};
\node (dec2) [decision, right of=dec1, xshift=2cm, yshift=-2cm] {
      \begin{tabular}{c}
      element  \\
      meet \\
      criteria 
      \end{tabular}
      };
\node (process1) [process, below of=dec2, yshift=-0.8cm] {signal server};
\node (process2) [process, right of=dec2, xshift=3cm] {move to next element};
\node (server) [textbox, right of=client, xshift=9.5cm] {\textbf{\underline{server}}};
\node (process3) [blocksignal, below of=server] {wait for signal};
\node  (process4) [process, below of=process3, xshift=1.8cm] {
\begin{tabular}{c}
      process  \\
      element 
      \end{tabular}
      };
\draw [arrow] (start) -- (dec1);
\draw [arrow] (dec1) -- node[anchor=south] {no} (dec2);
\draw [arrow] (dec1) -- node[anchor=east] {yes} (stop);
\draw [arrow] (dec2) -- node[anchor=south] {no} (process2);
\draw [arrow] (dec2) -- node[anchor=east] {yes} (process1);
\draw [arrow] (process3) -- (process4);
\draw [arrow] (process4) |- (process3);
\draw [dasharrow] (process1) -| (process3);
\draw [arrow] (process2) |- (dec1);

\end{tikzpicture}
\caption{Client server flow} \label{clientServer}
\end{figure}
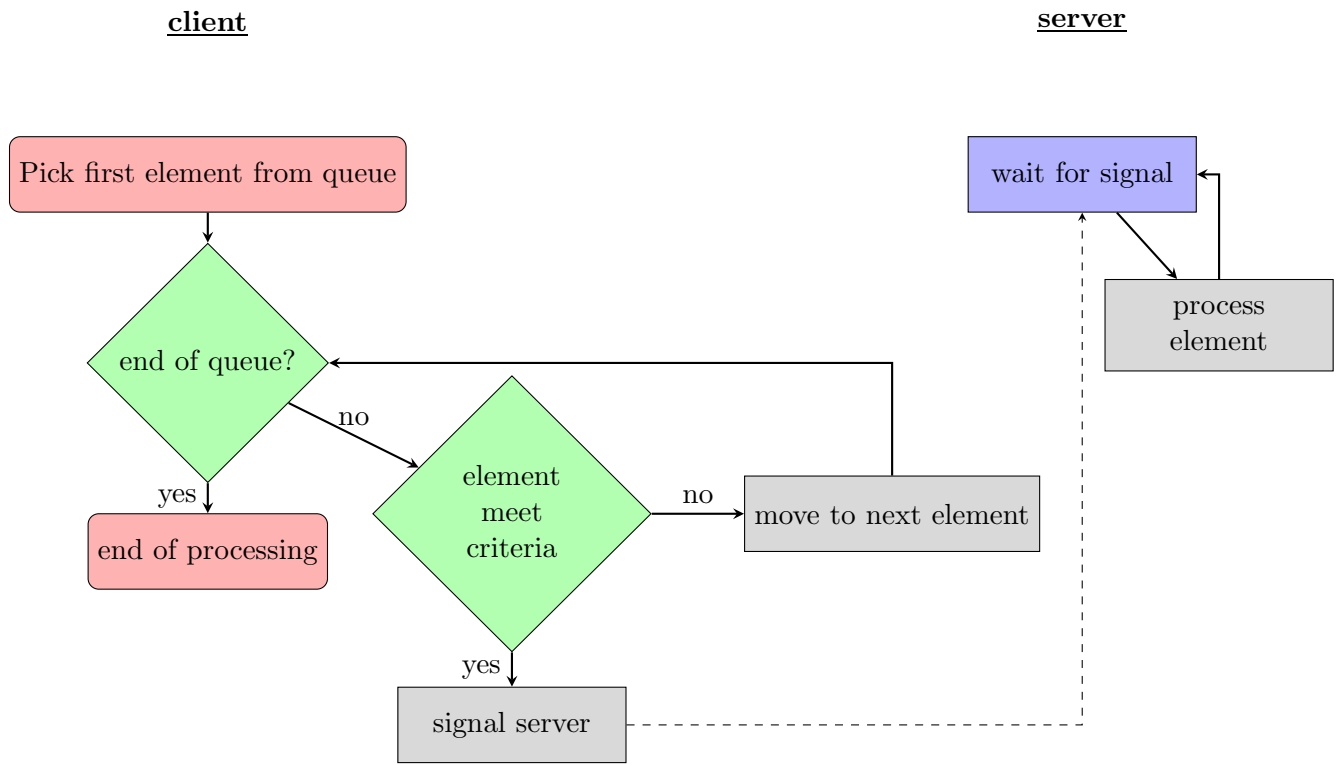

Suppose you have written some client/server code in which the client loops through a queue of data structures looking for one that meets some criteria.  When that criteria is met, it signals the server code to manipulate that data structure.  Furthermore, that data structure is serialized between the client and the server code (see \ref{clientServer})

As part of the code development process, the team needs to review the code.  Given the nature of the client/server code structure, the team decides to use the IRT (Interleaving Review Technique).  The main reason for using this code review technique is because IRT takes into account concurrency and fault tolerance.  For more information about IRT see \href{https://arxiv.org/abs/2407.02355}{Effective Technical Reviews - Chapter 4.6}

When attempting to review the system behavior of a concurrent/distributed and fault tolerant system, several problems arise including non-determinism and the state of the program.  Non-determinism because when the program is in some state, the next program state depends on which process executes next.  As a result, it is not clear how to proceed with the review process.  The state of the program is problematic because the current state is determined by all processes and their interrelated temporal dependencies. If we have three processes with 10 states each, we would have 1000 possible states to review.

One could focus the review process by having a reviewer take on the role of a devil's advocate.  This would be someone who is experienced in concurrent and fault tolerant systems.  Their role would be to make choices as to the timing of events and failures to maximize the probability that a bug is found during the review.  Deciding on these choices by hand is time consuming and error prone.

CTD can be used to aid the devil's advocate role in the code review.  The devil's advocate decides when another process/thread advances after or before a synchronization operation is performed.  They need to identify important information used at decision points to switch to different processes/thread and that information can be represented into attributes for a CTD model.  It is important to note that decision points should be made as concrete as possible.

The following example is a list of what could be important in making those decisions.
\begin{example}
\begin{itemize}
    \item Possible types of processes and threads
    \item Number of processes and threads
    \item Possible synchronization operations performed within a race window
    \item Possible shared resources being manipulated
    \item Types of manipulation on shared resources (read/write)
    \item Invoking system resources
    \item Possible values of shared variable
    \item Number of writers in the critical section that manipulates shared resources
    \item Restrictions (for example, can readers and writers be in the critical section that manipulates shared resources at the same time?)
\end{itemize}

\end{example}

The following example describes a sample code review.  Figure \ref{fig:codesnip1} is a pair of programs depicting client/server processing.  The relevant declares of variables are omitted.  G{\_} variables are common data structures that both the client and server have access to.  All others are local variables to either the client or server.  We assume that there is only one client and one server running at a time. G{\_}CBHead is the address of a CB structure. This is the head of a single-threaded queue or linked list.  G{\_}Num is the integer that is intended to represent the number of items that the server has to process.

\begin{figure}[hbt!]
    \begin{tabular}{ l l }
    \textbf{Client Pseudo Code Snippet} & \textbf{Server Pseudo Code Snippet} \\
 1. If (G{\_}Num == 0) then & 1. DO Forever;  \\ 
 2. DO; & 2. \hspace{.5cm}Wait for client signal;  \\  
 3. \hspace{.5cm}c{\_}Num = 0; & 3. \hspace{.5cm}DO s{\_}CBPtr=G{\_}CBHead \\  
 4. \hspace{.5cm}DO c{\_}CBPtr=G{\_}CBHead & 4.\hspace{1cm}Repeat(CB.Next) \\
 5. \hspace{1cm}Repeat(c{\_}CBPTR$->$CB.Next) & 5.\hspace{1cm}While(s{\_}CBPTR$<>$0); \\
 6. \hspace{1cm}While(c{\_}CBPtr$<>$ 0); & 6.\hspace{.7cm} If CB.Process $=$ ON Then \\
 7. \hspace{.7cm}If CB.Interesting Then & 7.\hspace{.7cm} DO; \\
 8. \hspace{.7cm}DO; & 8.\hspace{1cm} CB.Process$=$Off; \\
 9. \hspace{1cm}CB.Process$=$On; & 9.\hspace{1cm} Serialized(G{\_}Num $=$ G{\_}Num $-$1); \\
 10.\hspace{1cm}c{\_}Num $=$ c{\_}Num $+$ 1; & 10.\hspace{.7cm}END; \\
 11.\hspace{1cm}Signal Server; & 11.\hspace{.5cm}END; \\
 12.\hspace{.7cm}END; & 12. END; \\
 13.\hspace{.5cm} END; \\
 14.\hspace{.5cm}Serialized(G{\_}Num $=$ c{\_}Num); \\
 15. END;\\
 \\
 \\
 
    \end{tabular}

    \textbf{Variable Values:}
    \begin{itemize}
        \item c{\_}Num: 0
        \item G{\_}Num:0
        \item s{\_}CBPtr:0
        \item c{\_}CBPtr:0
    \end{itemize}
 
    \caption{Client/Server code snippets}
    \label{fig:codesnip1}
\end{figure}

The CB data type contains the following fields.
\begin{itemize}
  \item \textbf{Interesting} is a Boolean field that indicates whether the client found this CB "interesting". This may be randomly true or false.
  \item \textbf{Process} is also a Boolean field that indicates whether the server should process this CB.  This is set by the client.  We assume that this is false for all CBs to begin with.
  \item \textbf{Next} is the address of a CB structure.  This points to the next CB on the queue or linked list.
\end{itemize}

The local variables for the client are as follows.
\begin{itemize}
  \item \textbf{c{\_}CBPtr} is the address of a CB structure, and represents the current CB being processed by the client.  For brevity, any references to CB are actually using the dereferenced c{\_}CBPtr to locate the CB structure.  In other words, when we refer to CB.Process, we really mean c{\_}CBPtr$->$CB.Process.
  \item \textbf{C{\_}Num} is an integer and represents the updated number of the CBs for the server to process.
\end{itemize}

The server only has one local variable.
\begin{itemize}
  \item \textbf{s{\_}CBPtr} is the address of a CB structure, and represents the current CB being processed by the server.  Likewise, any references to CB use s{\_}CBPtr to locate the CB structure.
\end{itemize}

We also have the ability to perform serialized updates represented by Serialized(...).  This could be implemented by locks, semaphores, etc.

\break

In figure \ref{fig:codesnip2}, we assume the client code starts executing and there is only one CB data structure.  When it signals the server code (on line 11), the global count G{\_}Num has not been updated yet, but the CB data structure has been marked to be processed (on line 9).  Local variable c{\_}Num has been incremented to 1 and c{\_}CBPtr points to the first CB on the chain

\begin{figure}[hbt!]
    \begin{tabular}{ l l }
    \textbf{Client Pseudo Code Snippet} & \textbf{Server Pseudo Code Snippet} \\
 1. If (G{\_}Num == 0) then & 1. DO Forever;  \\ 
 2. DO; & 2. \hspace{.5cm}Wait for client signal;  \\  
 3. \hspace{.5cm}c{\_}Num = 0; & 3. \hspace{.5cm}DO s{\_}CBPtr=G{\_}CBHead \\  
 4. \hspace{.5cm}DO c{\_}CBPtr=G{\_}CBHead & 4.\hspace{1cm}Repeat(CB.Next) \\
 5. \hspace{1cm}Repeat(c{\_}CBPTR$->$CB.Next) & 5.\hspace{1cm}While(s{\_}CBPTR$<>$0); \\
 6. \hspace{1cm}While(c{\_}CBPtr$<>$ 0); & 6.\hspace{.7cm} If CB.Process $=$ ON Then \\
 7. \hspace{.7cm}If CB.Interesting Then & 7.\hspace{.7cm} DO; \\
 8. \hspace{.7cm}DO; & 8.\hspace{1cm} CB.Process$=$Off; \\
 9. \hspace{1cm}\textcolor{blue}{CB.Process$=$On;} & 9.\hspace{1cm} Serialized(G{\_}Num $=$ G{\_}Num $-$1); \\
 10.\hspace{1cm}c{\_}Num $=$ c{\_}Num $+$ 1; & 10.\hspace{.7cm}END; \\
 11.\hspace{1cm}\textcolor{blue}{Signal Server;} & 11.\hspace{.5cm}END; \\
 12.\hspace{.7cm}END; & 12. END; \\
 13.\hspace{.5cm} END; \\
 14.\hspace{.5cm}Serialized(G{\_}Num $=$ c{\_}Num); \\
 15. END;\\
 \\
 \\
 
    \end{tabular}

    \textbf{Variable Values:}
    \begin{itemize}
        \item \textcolor{blue}{c{\_}Num: \cancel{\textcolor{red}0} \textbf{1}}
        \item G{\_}Num:0
        \item s{\_}CBPtr:0
        \item \textcolor{blue}{c{\_}CBPtr:\cancel{\textcolor{red}0} \textbf{First CB}}
    \end{itemize}
 
    \caption{Client/Server code snippets}
    \label{fig:codesnip2}
\end{figure}

\break
Once the server has been signaled, in figure \ref{fig:codesnip3}, we switch processing to the server, which can wake up from the wait (on line 2) and process the CB data structure that has been marked.  The server can then decrement the global count variable G{\_}Num (on line 9) from 0 to -1.  The Server will now exit the s{\_}CBPtr loop and return to waiting for a signal (line 2).
\begin{figure}[hbt!]
    \begin{tabular}{ l l }
    \textbf{Client Pseudo Code Snippet} & \textbf{Server Pseudo Code Snippet} \\
 1. If (G{\_}Num == 0) then & 1. DO Forever;  \\ 
 2. DO; & 2. \hspace{.5cm}\textcolor{blue}{Wait for client signal;}  \\  
 3. \hspace{.5cm}c{\_}Num = 0; & 3. \hspace{.5cm}DO s{\_}CBPtr=G{\_}CBHead \\  
 4. \hspace{.5cm}DO c{\_}CBPtr=G{\_}CBHead & 4.\hspace{1cm}Repeat(CB.Next) \\
 5. \hspace{1cm}Repeat(c{\_}CBPTR$->$CB.Next) & 5.\hspace{1cm}While(s{\_}CBPTR$<>$0); \\
 6. \hspace{1cm}While(c{\_}CBPtr$<>$ 0); & 6.\hspace{.7cm} If CB.Process $=$ ON Then \\
 7. \hspace{.7cm}If CB.Interesting Then & 7.\hspace{.7cm} DO; \\
 8. \hspace{.7cm}DO; & 8.\hspace{1cm} CB.Process$=$Off; \\
 9. \hspace{1cm}CB.Process$=$On; & 9.\hspace{1cm} \textcolor{blue}{Serialized(G{\_}Num $=$ G{\_}Num $-$1);} \\
 10.\hspace{1cm}c{\_}Num $=$ c{\_}Num $+$ 1; & 10.\hspace{.7cm}END; \\
 11.\hspace{1cm}Signal Server; & 11.\hspace{.5cm}END; \\
 12.\hspace{.7cm}END; & 12. END; \\
 13.\hspace{.5cm} END; \\
 14.\hspace{.5cm}Serialized(G{\_}Num $=$ c{\_}Num); \\
 15. END;\\
 \\
 \\
 
    \end{tabular}

    \textbf{Variable Values:}
    \begin{itemize}
        \item c{\_}Num: 1
        \item \textcolor{blue}{G{\_}Num: \cancel{\textcolor{red}0} \textbf{-1}}
        \item \textcolor{blue}{s{\_}CBPtr:\cancel{\textcolor{red}0} \textbf{First CB}}
        \item c{\_}CBPtr: First CB
    \end{itemize}
 
    \caption{Client/Server code snippets}
    \label{fig:codesnip3}
\end{figure}

\newpage
Next in figure \ref{fig:codesnip4}, we return to the client, which attempts to update G{\_}Num (line 14), which happens after the Server code runs and updates G{\_}Num.  The client now exits.

\begin{figure}[hbt!]
    \begin{tabular}{ l l }
    \textbf{Client Pseudo Code Snippet} & \textbf{Server Pseudo Code Snippet} \\
 1. If (G{\_}Num == 0) then & 1. DO Forever;  \\ 
 2. DO; & 2. \hspace{.5cm}Wait for client signal;  \\  
 3. \hspace{.5cm}c{\_}Num = 0; & 3. \hspace{.5cm}DO s{\_}CBPtr=G{\_}CBHead \\  
 4. \hspace{.5cm}DO c{\_}CBPtr=G{\_}CBHead & 4.\hspace{1cm}Repeat(CB.Next) \\
 5. \hspace{1cm}Repeat(c{\_}CBPTR$->$CB.Next) & 5.\hspace{1cm}While(s{\_}CBPTR$<>$0); \\
 6. \hspace{1cm}While(c{\_}CBPtr$<>$ 0); & 6.\hspace{.7cm} If CB.Process $=$ ON Then \\
 7. \hspace{.7cm}If CB.Interesting Then & 7.\hspace{.7cm} DO; \\
 8. \hspace{.7cm}DO; & 8.\hspace{1cm} CB.Process$=$Off; \\
 9. \hspace{1cm}CB.Process$=$On; & 9.\hspace{1cm} Serialized(G{\_}Num $=$ G{\_}Num $-$1); \\
 10.\hspace{1cm}c{\_}Num $=$ c{\_}Num $+$ 1; & 10.\hspace{.7cm}END; \\
 11.\hspace{1cm}Signal Server; & 11.\hspace{.5cm}END; \\
 12.\hspace{.7cm}END; & 12. END; \\
 13.\hspace{.5cm} END; \\
 14.\hspace{.5cm}\textcolor{blue}{Serialized(G{\_}Num $=$ c{\_}Num);} \\
 15. END;\\
 \\
 \\
 
    \end{tabular}

    \textbf{Variable Values:}
    \begin{itemize}
        \item c{\_}Num: 1
        \item \textcolor{blue}{G{\_}Num: \cancel{\textcolor{red}-1} \textbf{1}}
        \item s{\_}CBPtr:First CB
        \item c{\_}CBPtr: First CB
    \end{itemize}
 
    \caption{Client/Server code snippets}
    \label{fig:codesnip4}
\end{figure}

\newpage
Any future instance of the Client, in \ref{fig:codesnip5}, will see that G{\_}Num is non-zero (line 1) and skip any processing that will signal the server code (line 15).  Likewise, the Server will not get control to update the count since it is waiting for a signal that will never occur from the Client.

\begin{figure}[hbt!]
    \begin{tabular}{ l l }
    \textbf{Client Pseudo Code Snippet} & \textbf{Server Pseudo Code Snippet} \\
 1. \textcolor{blue}{If (G{\_}Num == 0) then} & 1. DO Forever;  \\ 
 2. DO; & 2. \hspace{.5cm}Wait for client signal;  \\  
 3. \hspace{.5cm}c{\_}Num = 0; & 3. \hspace{.5cm}DO s{\_}CBPtr=G{\_}CBHead \\  
 4. \hspace{.5cm}DO c{\_}CBPtr=G{\_}CBHead & 4.\hspace{1cm}Repeat(CB.Next) \\
 5. \hspace{1cm}Repeat(c{\_}CBPTR$->$CB.Next) & 5.\hspace{1cm}While(s{\_}CBPTR$<>$0); \\
 6. \hspace{1cm}While(c{\_}CBPtr$<>$ 0); & 6.\hspace{.7cm} If CB.Process $=$ ON Then \\
 7. \hspace{.7cm}If CB.Interesting Then & 7.\hspace{.7cm} DO; \\
 8. \hspace{.7cm}DO; & 8.\hspace{1cm} CB.Process$=$Off; \\
 9. \hspace{1cm}CB.Process$=$On; & 9.\hspace{1cm} Serialized(G{\_}Num $=$ G{\_}Num $-$1); \\
 10.\hspace{1cm}c{\_}Num $=$ c{\_}Num $+$ 1; & 10.\hspace{.7cm}END; \\
 11.\hspace{1cm}Signal Server; & 11.\hspace{.5cm}END; \\
 12.\hspace{.7cm}END; & 12. END; \\
 13.\hspace{.5cm} END; \\
 14.\hspace{.5cm}Serialized(G{\_}Num $=$ c{\_}Num); \\
 15. \textcolor{blue}{END}\\
 \\
 \\
 
    \end{tabular}

    \textbf{Variable Values:}
    \begin{itemize}
        \item c{\_}Num: 1
        \item G{\_}Num: 1
        \item s{\_}CBPtr:First CB
        \item c{\_}CBPtr: First CB
    \end{itemize}
 
    \caption{Client/Server code snippets}
    \label{fig:codesnip5}
\end{figure}

There are many potential points in the previous example to switch processes.  CTD can be used to build a list of interesting scenarios to try out during the review process.  For instance look at example \ref{exp:ctdreview}.
\begin{example}
    \textbf{Multiple data structures (CB) elements on a chain.}
    \begin{itemize}
      \item Small, medium, large number of elements
    \end{itemize}  
    \textbf{Data structure (CB) elements that are interesting.}
    \begin{itemize}
      \item One or multiple are interesting
      \item Where on the chain? First; Middle; Last
    \end{itemize}  
    \textbf{Number of client processes running at the same time}
    \begin{itemize}
      \item Small, medium, large number of processes
    \end{itemize}  
     \label{exp:ctdreview}
\end{example}

Starting with a very simple CTD model, there are two obvious inputs to our Client code snippet.  The length of the chain of CBs, and which CBs are interesting to be marked as processed.  Following is example \ref{exp:codemodel} of what this model would look like.
\begin{example}
The number of CB's on the chain could be very small or very large.  For the purposes of building a model, let's use 0 to 5 elements.
    \begin{itemize}
      \item LenCBchain=$<0, 1, 2, 3, 4, 5>$
    \end{itemize}  
Which CB's are interesting?  It could be all, none, or a mix.  For simplicity in building the model, we will treat a non-existent CB the same as not interesting.  If there is only one CB, we will treat the interesting flag as false for the second and subsequent CBs.
    \begin{itemize}
      \item InterestingCB1=$<true,false>$
      \item InterestingCB2=$<true,false>$
      \item InterestingCB3=$<true,false>$
      \item InterestingCB4=$<true,false>$
      \item InterestingCB5=$<true,false>$
    \end{itemize}  
    \label{exp:codemodel}
\end{example}
\newpage
This gives us a list of 63 legal combinations.  This is way too many to work with.  Using CTD to reduce the number of tests can get us to a manageable 13 combinations in Table \ref{tab:codeReduction}.  We used line 5 in the earlier example \ref{fig:codesnip2}.

\begin{table}[]
    \begin{tabular}{l|l|l|l|l|l|l}
\textbf{\tiny{Index}} & \textbf{\tiny{LengthOfChain}} &\textbf{\tiny{InterestingCB1}} &\textbf{\tiny{InterestingCB2}} &\textbf{\tiny{InterestingCB3}} &\textbf{\tiny{InterestingCB4}} &\textbf{\tiny{InterestingCB5}} \\

1 & 3 & true & true & true & false & false  \\
2 & 4 & false & false & false & true & false  \\
3 & 5 & true & true & false & true & true  \\
4 & 2 & false & true & false & false & false  \\
\cellcolor{yellow}{5} & \cellcolor{yellow}{1} & \cellcolor{yellow}{true} &\cellcolor{yellow}{false} &\cellcolor{yellow}{false} &\cellcolor{yellow}{false} &\cellcolor{yellow}{false} \\
6 & 5 & false & false & true & false & false  \\
7 & 0 & false & false & false & false & false  \\
8 & 4 & true & true & true & false & false  \\
9 & 3 & false & false & false & false & false  \\
10 & 2 & true & false & false & false & false  \\
11 & 1 & false & false & false & false & false  \\
12 & 5 & false & false & true & true & true  \\
13 & 5 & false & true & false & false & true 
    \end{tabular}
    \caption{Test Reduction for Code Review example}
    \label{tab:codeReduction}
\end{table}

To take this example a step further, we can refine the model.  What if we were to add to the model the point when we want to dispatch the server task?  Assume the same dispatch point (immediately after the signal).  Pick which element in the chain is being processed when we dispatch the server process.

\begin{itemize}
    \item DispatchAfter (DA) = $<1, 2, 3, 4, 5>$
    \item Restrictions: DA can not be greater than LengthOfChain $+1$
\end{itemize}

That can get us to 130 possible combinations (assuming we also add some sensible restrictions.)  Using CTD test reduction, we can get it down to 19 tests see table  \ref{tab:codeReductionDispatch}.  Note that line 1 was used in our example.
\newpage
\begin{table}[]
    \begin{tabular}{l|l|l|l|l|l|l|l}
\textbf{\tiny{Index}} & \textbf{\tiny{LengthOfChain}} &\textbf{\tiny{InterestingCB1}} &\textbf{\tiny{InterestingCB2}} &\textbf{\tiny{InterestingCB3}} &\textbf{\tiny{InterestingCB4}} &\textbf{\tiny{InterestingCB5}} &\textbf{\tiny{DA}} \\

\cellcolor{yellow}{1} & \cellcolor{yellow}{1} & \cellcolor{yellow}{true} & \cellcolor{yellow}{false} & \cellcolor{yellow}{false} & \cellcolor{yellow}{false} & \cellcolor{yellow}{false} & \cellcolor{yellow}{1}  \\
2 & 4 & false & true & true & true & false & 4  \\
3 & 5 & true & true & true & true & true & 3  \\
4 & 3 & false & true & false & false & false & 2  \\
5 & 0 & false & false & false & false & false & 0 \\
6 & 5 & false & false & false & false & true & 5  \\
7 & 3 & false & false & true & false & false & 3 \\
8 & 4 & true & true & true & true & false & 1  \\
9 & 2 & true & true & false & false & false & 2  \\
10 & 5 & true & false & false & true & true & 4  \\
11 & 5 & true & true & true & true & true & 2  \\
12 & 3 & true & false & true & false & false & 1  \\
13 & 4 & true & false & true & false & false & 3 \\
14 & 2 & true & false & false & false & false & 1 \\
15 & 5 & true & true & true & true & true & 1 \\
16 & 4 & true & true & false & true & false & 2 \\
17 & 2 & false & true & false & false & false & 2 \\
18 & 5 & true & true & true & true & true & 5 \\
19 & 5 & false & true & true & false & false & 3 \\
    \end{tabular}
    \caption{Test Reduction for Code Review example with Dispatch attribute}
    \label{tab:codeReductionDispatch}
\end{table}

In our example \ref{tab:codeReductionDispatch}, even 19 rows is probably too many to cover in a code review.  Extreme cases, such as line 1 (1 CB) and lines 3 or 11 (5 CBs) are good choices.  "Middle of the road" cases are also good choices, such as line 16, which has 4 CMs and we choose to dispatch in the middle of the chain (see \ref{tab:codeReductionDispatchExtreme})
\newpage
\begin{table}[]
    \begin{tabular}{l|l|l|l|l|l|l|l}
\textbf{\tiny{Index}} & \textbf{\tiny{LengthOfChain}} &\textbf{\tiny{InterestingCB1}} &\textbf{\tiny{InterestingCB2}} &\textbf{\tiny{InterestingCB3}} &\textbf{\tiny{InterestingCB4}} &\textbf{\tiny{InterestingCB5}} &\textbf{\tiny{DA}} \\

\cellcolor{yellow}{1} & \cellcolor{yellow}{1} & \cellcolor{yellow}{true} & \cellcolor{yellow}{false} & \cellcolor{yellow}{false} & \cellcolor{yellow}{false} & \cellcolor{yellow}{false} & \cellcolor{yellow}{1}  \\
2 & 4 & false & true & true & true & false & 4  \\
\cellcolor{yellow}{3} & \cellcolor{yellow}{5} & \cellcolor{yellow}{true} & \cellcolor{yellow}{true} & \cellcolor{yellow}{true} & \cellcolor{yellow}{true} & \cellcolor{yellow}{true} & \cellcolor{yellow}{3}  \\
4 & 3 & false & true & false & false & false & 2  \\
5 & 0 & false & false & false & false & false & 0 \\
6 & 5 & false & false & false & false & true & 5  \\
7 & 3 & false & false & true & false & false & 3 \\
8 & 4 & true & true & true & true & false & 1  \\
9 & 2 & true & true & false & false & false & 2  \\
10 & 5 & true & false & false & true & true & 4  \\
\cellcolor{yellow}{11} & \cellcolor{yellow}{5} & \cellcolor{yellow}{true} & \cellcolor{yellow}{true} &\cellcolor{yellow}{true} & \cellcolor{yellow}{true} & \cellcolor{yellow}{true} & \cellcolor{yellow}{2}  \\
12 & 3 & true & false & true & false & false & 1  \\
13 & 4 & true & false & true & false & false & 3 \\
14 & 2 & true & false & false & false & false & 1 \\
15 & 5 & true & true & true & true & true & 1 \\
\cellcolor{yellow}{16} & \cellcolor{yellow}{4} & \cellcolor{yellow}{true} & \cellcolor{yellow}{true} & \cellcolor{yellow}{false} & \cellcolor{yellow}{true} & \cellcolor{yellow}{false} & \cellcolor{yellow}{2} \\
17 & 2 & false & true & false & false & false & 2 \\
18 & 5 & true & true & true & true & true & 5 \\
19 & 5 & false & true & true & false & false & 3 \\
    \end{tabular}
    \caption{Test Reduction for Code Review example with Dispatch attribute}
    \label{tab:codeReductionDispatchExtreme}
\end{table}

Creating a model for use in a code review is a different exercise than using it for testing.  Here are a few pitfalls to avoid.
\begin{itemize}
    \item Do not spend an excessive amout of time creating a model.  No more than 5-10 minutes.
    \item Creating a complicated model with lots of restrictions can be very time consuming.  Build a few smaller models instead.
    \item Although our examples have focused on modeling the state of the system, modeling the inputs to a segment of code can also have value.
    \item After the code review, the model can be enhanced for use in testing or long-term use.
\end{itemize}

%\chapter{Out of the box applications of CTD}

%TBC it is just a cartesian product.  Example of out of the box applications

%\chapter{Relation of CTD to experimental design and Machine Learning}

%When conducting a scientific experiements factors that impact the indepdent variable are idenfied.  Thus, a Cartesian product is created that requires to many experiements.  CTD is a way to overcome this but further assumption are added to the relation between the indendent and dependent factors.

\appendix

\chapter{Exercises}

In what follows a set of exercises are introduced that will enhance your understanding of the combinatorial testing technique.   These exercises are intentionally at the concept level and their solution does not require any programming.

\begin{exercise} \label{Exercise:Bank1}
An ATM machine is used to withdraw money by the bank’s customers.   It is connected to the bank central DB through a network but the connection may fail.  Only if a withdrawal is above $\$20$ the bank’s central DB is accessed to check if there is enough money in the customer account.   How many combinations are at least applied by this description – 
\begin{enumerate}
    \item Two
    \item Four
    \item Three
\end{enumerate}
See solution \ref{Exercise:Bank1:solution}
\end{exercise}

\begin{exercise} \label{Exercise:Bank2}
Enumerate the combinations appearing in exercise \ref{Exercise:Bank1}
See solution \ref{Exercise:Bank2:solution}
\end{exercise}

\begin{exercise} \label{Exercise:Bank3}
State the questions you would ask to clarify the description in exercise \ref{Exercise:Bank1}
See solution \ref{Exercise:Bank3:solution}
\end{exercise}

\begin{exercise} \label{Exercise:Bank4}
The following information was provided to enhance the Exercise \ref{Exercise:Bank1}: Smaller than $\$20$ withdrawals can occur only once without updating the bank central DB.  How many combinations do we have now?
\begin{enumerate}
    \item four
    \item six
    \item eight
\end{enumerate}
\end{exercise}

\begin{exercise} \label{Exercise:Bank5}
Enumerate the combinations appearing in exercise \ref{Exercise:Bank4}
\end{exercise}

\begin{exercise} \label{Exercise:Bank6}
What are the above tests clearly missing in exercise \ref{Exercise:Bank5}
\begin{enumerate}
    \item Expected results 
    \item Concrete values for the withdrawal amount 
    \item Type of network failures
    \item Options one and two
\end{enumerate}
See solution \ref{Exercise:Bank6:solution}
\end{exercise}

\begin{exercise} \label{Exercise:Cartesian1}
Given 3 sets of attributes each with 3 values, the Cartesian product is
\begin{enumerate}
    \item eight
    \item nine
    \item six
\end{enumerate}
See solution \ref{Exercise:Cartesian1:solution}
\end{exercise}

\begin{exercise} \label{Exercise:Cartesian2}
Given n sets of attributes each with k values, the Cartesian product is
\begin{enumerate}
    \item n * k
    \item $(k + ... +k)$ n times
    \item $(K * ... *k)$ n times
\end{enumerate}
See solution \ref{Exercise:Cartesian2:solution}
\end{exercise}

\begin{exercise} \label{Exercise:Cartesian3}
Given $X ={1,2,3}$ and $Y={a,b}$ the Cartesian product X*Y is
\begin{enumerate}
    \item ${(1,z),(2,a),(3,a),(1,b),(2,b),(3,b)}$
    \item ${1,2,3,a,b}$
    \item ${(1,a),2,a),(3,a),(1,b),2,b),(3,b),(1,c),(2,c),3,c)}$
\end{enumerate}
See solution \ref{Exercise:Cartesian3:solution}
\end{exercise}

\begin{exercise} \label{Exercise:Cartesian4}
If you take away the pairs that end with “a” in X*Y of Exercise \ref{Exercise:Cartesian3}
\begin{enumerate}
    \item $(1, b), (2, b), (3, b)$
    \item ${1, 2, 3, b}$
    \item {(1, b), (2, b), (3, b), (1, c), (2, c), (3, c)}
\end{enumerate}

This set can be defined as all the elements $(x, y)$ in X*Y such that $y <> a$ (this is referred to as a constraint)
See solution \ref{Exercise:Cartesian4:solution}
\end{exercise}

\begin{exercise} \label{Exercise:Shopping1}
Suppose we have an Online Shopping System with the following parameters
\begin{enumerate}
    \item Availability
    \item Payment method
    \item Carrier
    \item Delivery schedule
    \item Export control
\end{enumerate}

The table \ref{table:Online Shopping System} represents the possible values for each of the parameters. A test is represented by an assignment of exactly one value to each parameter.  In this Shopping System example, we have $4x3x3x4x2 = 288$ combinations.

\begin{table}
    \centering
    \begin{tabular}{|c|c|c|c|c|} \hline 
         \textbf{Availability}&  \textbf{Payment}&  \textbf{Carrier}&  \textbf{Delivery schedule}& \textbf{Export control}\\ \hline 
         Available&  Credit&  Mail&  One Day& True\\ \hline 
         Not in Stock&  Paypal&  UPS&  2-5 working days& False\\ \hline 
         Discontinued&  Gift Voucher&  Fedex&  6-10 working days& \\ \hline 
         No Such Product&  &  &  Over 10 working days& \\ \hline
    \end{tabular}
    \caption{Online Shopping system}
    \label{table:Online Shopping System}
\end{table}

Based on the above Online Shopping System, How would you choose tests for this system?

See solution \ref{Exercise:Shopping1:solution}
\end{exercise}

\begin{exercise} \label{Exercise:Shopping2}
Based on the above Online Shopping System in Exercise \ref{Exercise:Shopping1}, How many tests do you require?
See solution \ref{Exercise:Shopping2:solution}    
\end{exercise}

\begin{exercise} \label{Exercise:Shopping3}
Based on the above Online Shopping System in Exercise \ref{Exercise:Shopping1}, How do you review your choices?
See solution \ref{Exercise:Shopping3:solution}    
\end{exercise}

\begin{exercise} \label{Exercise:Shopping4}
Based on the above Online Shopping System in Exercise \ref{Exercise:Shopping1}, How do you prove the validity of your choices? 
\end{exercise}

\begin{exercise} \label{Exercise:Shopping5}
Based on the above Online Shopping System in Exercise \ref{Exercise:Shopping1}, Suppose there is a bug and \textbf{Credit} does not work well with \textbf{One Day} delivery.  Any combination that includes \textbf{Credit} and a \textbf{One Day} delivery will expose that bug.  There are 24 such combinations (All combinations in which payment=\textbf{Credit} AND delivery=\textbf{One Day}.  Values for availability, carrier and export control are free.  This would give us $4x3x2=24$.

Suppose \textbf{Credit} does not work well with a \textbf{One Day} delivery, but only with \textbf{Fedex}.  Any combination that includes \textbf{Credit}, \textbf{One Day} delivery, and \textbf{Fedex} will expose that bug.  There will be 8 such combinations

We will call the first case a \textbf{level two interaction} and the second case a \textbf{level three interaction}

Explain the 8 combinations for \textbf{Credit}, \textbf{One Day} and \textbf{Fedex}
See solution \ref{Exercise:Shopping5:solution}    
\end{exercise}

\begin{exercise} \label{Exercise:Shopping6}
Based on the above Online Shopping System in Exercise \ref{Exercise:Shopping5}, How many combinations will expose a bug with Delivery schedule=\textbf{One Day} AND Export control=\textbf{True}?
See solution \ref{Exercise:Shopping6:solution}    
\end{exercise}

\newpage
\begin{exercise} \label{Exercise:Interaction1}
The root cause analysis of many bugs shows they depend on a value of one variable $(20\%-68\%)$.  Most defects can be discovered in tests of interactions between the values of two variables $(65-97\%)$

\begin{table}[hbt!]
    \centering
    \begin{tabular}{|c|c|c|c|c|c|} \hline 
         \textbf{Vars}&  \textbf{Medical Devices}& \textbf{Browser}&  \textbf{Server}&  \textbf{NASA GSFC}& \textbf{Network Security}\\ \hline 
         1&  66&  29&  42&  68& 20\\ \hline 
         2&  97&  76&  70&  93& 65\\ \hline 
         3&  99&  95&  89&  98& 90\\ \hline 
         4&  100&  97&  96&  100& 98\\ \hline 
         5&  &  99&  96&  & 100\\ \hline 
 6& & 100& 100& &\\ \hline
    \end{tabular}
    \caption{Number of variables involved in triggering software faults\\
    Source http://csrc.nist.gov/groups/SNS/acts/ftfi.html}
    \label{table:interaction1}
\end{table}

If you are in network security and want to get $95\%$ confidence, using the Table \ref{table:interaction1} what level of interaction do you require?
See solution \ref{Exercise:Interaction1:solution}    
\end{exercise}

\begin{exercise}  \label{Exercise:Interaction2}
If you are in some general domain and want $95\%$ confidence, using the Table \ref{table:interaction1} what level of interaction do you require?
See solution \ref{Exercise:Interaction2:solution}     
\end{exercise}

\begin{exercise}  \label{Exercise:Interaction3}
Using the Online Shopping System defined in Exercise \ref{Exercise:Shopping1} a level 2 interaction level would result in 101 different pairs of values. 
\begin{example}
    \begin{itemize}
        \item Payment=Credit, Delivery=One Day
        \item Payment=Credit, Delivery= 2-5 Days
        \item ...
        \item Availability=Available, Delivery=One Day
        \item ...
    \end{itemize}
\end{example}
A given test plan covers x\% of interactions level 2 if it covers x\% of these 101 pairs.  100\% pairwise coverage means that each pair appears at least once.  A test plan that gives 100\% pairwise coverage will reveal \textbf{all} defects that result from an interaction level of 2.

To explain the \textbf{101} number above we would sum the following
\begin{itemize}
    \item 3x4=12 pairs for Payment AND Delivery
    \item 4x4=16 pairs for Availability AND Deliver
    \item etc.
\end{itemize}

How many triplets are there in level 3 interaction?
See solution \ref{Exercise:Interaction3:solution}     
\end{exercise}

\begin{exercise}  \label{Exercise:Interaction4}
If we have $X={a,b}, Y={c,d}, Z={e,f}$, how many interaction level two do you have in X*Y*Z?
\begin{enumerate}
    \item twelve
    \item four
    \item eight
\end{enumerate}
Enumerate them
See solution \ref{Exercise:Interaction4:solution}  
\end{exercise}

\begin{exercise}  \label{Exercise:Interaction5}
If we have $X={a,b}, Y={c,d}, Z={e,f}$, and we have the \textbf{a} dropped from all triples, how many interaction level two are we left with?
\begin{enumerate}
    \item ten
    \item eight
    \item six   
\end{enumerate}
Enumerate them
See solution \ref{Exercise:Interaction5:solution}  
\end{exercise}

\begin{exercise}  \label{Exercise:Interaction6}
Explain how a model can test for all interaction level two but still miss a bug on interaction level two of the given model?
See solution \ref{Exercise:Interaction6:solution}  
\end{exercise}

\begin{exercise}  \label{Exercise:Interaction7}
How many (and which) pairs are covered by the following combination:
(Availability=available, Payment=paypal, Carrier=fedEx, Schedule=$2-5$working days, ExportControl=True).  Explain how to get the number without counting.
See solution \ref{Exercise:Interaction7:solution}  
\end{exercise}

\begin{exercise}  \label{Exercise:Interaction8}
Is there a difference between the following two requirements where the coverage goal is every 2 attributes?
\begin{enumerate}
    \item Attributes: Availability, Payment, Carrier, Delivery Schedule, Export Control
    \item Attributes: Availability, Payment
\end{enumerate}
See solution \ref{Exercise:Interaction8:solution}  
\end{exercise}

\begin{exercise}  \label{Exercise:Interaction9}
In the above exercise \ref{Exercise:Interaction8}, enumerate the pair values that are guaranteed by the first coverage requirement but not guaranteed by the second.  How many coverage requirement pairs did you get?
See solution \ref{Exercise:Interaction9:solution}  
\end{exercise}

\begin{exercise}  \label{Exercise:Communication1}
We would like to test a group communication protocol.  Nodes can be either up or down and for each pair of nodes communication can be either up or down.  The current group communication is correct if it is a maximal group with all its nodes up and if a pair of them are connected.

Model the different possible failure scenarios for three nodes
\end{exercise}

\begin{exercise}  \label{Exercise:Open1}
Create a model to test the open() Linux system call based on the description in http://linux.die.net/man/2/open.  Come up with the initial model, but state explicitly what part of the description of the API you are not modeling yet.
This exercise may take some time to fully complete as it is more of a project than an exercise.  
\end{exercise}

% DEBBIE to continue here

\begin{exercise}  \label{Exercise:AddingAttributes}
Suppose we have the following model
\begin{itemize}
    \item Attribute1 with value1
    \item Attribute2 with value1, value2
    \item Attribute3 with value1, value2, value3
    \item Attribute4 with value1, value2, value3, value4
    \item Attribute5 with value1, value2, value3, value4, value5
\end{itemize}

The complete Cartesian product of the above model is $1 \times 2 \times 3 \times 4 \times 5 = 120$ tests.  

Clearly, the number of tests grows quickly when adding more and more values.  In chapter \ref{ChapterCoverageReduction}, we introduced pair coverage and the reduction obtained by designing a minimal test sets that is a subset of the 120 tests in the above Cartesian product that achieves pair coverage.  Assuming we obtain such a set of tests one may expect the number of tests to reduce to on the order of 20 tests. 

The reader can go ahead and try to obtain the above reduction using one of the available free 
\href{https://www.geeksforgeeks.org/combinatorial-testing-tools-in-software-testing/}{combinatorial test design tools.}

\vspace{5mm} %5mm vertical space
Given the set of tests you have designed that achieve pair coverage, which attributes in the above model can have additional values added to them without increasing the number of reduced tests?

See solution \ref{Exercise:AddingAttributes:solution}
\end{exercise}

\begin{exercise}  \label{Exercise:AddingAttributes2}
Suppose we have the following model
\begin{itemize}
    \item Attribute1 with value1, value2, value3, value4, value5
    \item Attribute2 with value1, value2, value3, value4, value5
    \item Attribute3 with value1, value2, value3, value4, value5
    \item Attribute4 with value1, value2, value3, value4, value5
    \item Attribute5 with value1, value2, value3, value4, value5
\end{itemize}

The complete Cartesian product of the above model is $5 \times 5 \times 5 \times 5 \times 5 = 3125$ tests.   When running the CTD pair-wise test reduction, one may expect the number of tests to reduce to on the order of 32 tests.

\vspace{5mm} %5mm vertical space
Given the set of tests you have designed that achieve pair coverage, which attributes in the above model can have additional values added to them without increasing the number of reduced tests?
 
See solution \ref{Exercise:AddingAttributes2:solution}
\end{exercise}

\begin{exercise} \label{Exercise:FunctionCall}
In this exercise we are looking at a simple function call

\textbf{void foo(int level, Boolean isBar, String menuSelection) {}}
\begin{itemize}
      \item level with values $0-9$
      \item isBar with values true, false
      \item menuSelection with values "status", "list", "add", "remove"
\end{itemize}  
If we were to perform a unit test on this function call, we would like to cover each parameter at least once.  To do this we would total the sum of each possible value for each attribute $10 level values + 2 isBar values + 4 menuSelection values = 16 tests$

\vspace{5mm} %5mm vertical space
However, if we wanted to perform a functional test, we would want to cover every interaction.  

Given the above function call and possible values, how many tests would be in the complete Cartesian product?
See solution \ref{Exercise:FunctionCall:solution}
\end{exercise}
 
\begin{exercise} \label{Exercise:manualCTD}
Consider a function with three-parameters, each of which can take three values
\begin{itemize}
    \item Color: red, white, blue
    \item Size: small, medium, large
    \item Quantity: 1, 2, 3
\end{itemize}
The complete Cartesean product would be %3 * 3 * 3 = 27 tests%.
Construct sets of tests that maximize efficiency by duplicating sets as little as possible. (manually perform the pair-wise reduction)
See solution \ref{Exercise:manualCTD:solution}
\end{exercise}

\begin{exercise} \label{Exercise:pairwiseNum}
How much reduction can occur?  It depends upon the shape of the input matrix and the interaction level chosen.

For example: 3 attributes with 3 values each.  Complete Cartesian product is 27 tests.  With pairwise can reduce to 9 tests.

For 8 attributes with 2 values each.
\begin{enumerate}
    \item What is the complete Cartesian product? 
    \item What is the pairwise reduction? 
\end{enumerate}

See solution \ref{Exercise:pairwiseNum:solution}
\end{exercise}

% Commenting out chapters that still need work but that will not be part of the first draft
%\chapter{Use of CTD in design and requirements specification}
%use of CTD on top of use cases using it to go from user stories to use cases
%\chapter{Advance application of CTD modeling}
%TBC testing of sequences,
%\chapter{Automation, abstract to concrete, expected results}

\chapter{CTD resources and related work}

Here we will provide an annotated list of potential further reading on CTD as well as point to resources that could be used to implement the approach. 

A list of available CTD tools is provided here 
\href{https://www.geeksforgeeks.org/combinatorial-testing-tools-in-software-testing/}{link}.

\chapter{Solutions to Exercises}

\section{Exercise \ref{Exercise:SourceInfo}}
\label{Exercise:SourceInfo:solution}
The best answer is option 4

\section{Exercise \ref{Exercise:AddedFunction}}
\label{Exercise:AddedFunction:solution}
The best answer is options 1 and 3.  If not concerned about the remote communication then randomizing at the testing level is the best answer.  If concerned about mistakes in the implementation, then adding the attribute to the model is the best answer.

\section{Exercise \ref{Exercise:Constraint1}}
\label{Exercise:Constraint1:solution}
The best answer is option 4.  Removing values from Attribute1 first will reduce the Cartesian product because Attribute1 has the largest number of values.  Removing values from largest to smallest will produce the greatest yield of reduced number of tests.

\section{Exercise \ref{Exercise:Bank1}}
\label{Exercise:Bank1:solution}
The answer is option $3$ as going to the central bank DB when the withdrawal is less than $\$20$ is not possible. 

\section{Exercise \ref{Exercise:Bank2}}
\label{Exercise:Bank2:solution}
The network$\_$connection can succeed or fail.  The customer's bank account can have enough money or not enough money.  So, the enumeration would be as follows
\begin{itemize}
    \item network$\_$connection $=$ succeed AND bank$\_$account $=$ enough
    \item network$\_$connection $=$ succeed AND bank$\_$account $=$ not$\_$enough
    \item network$\_$connection $=$ fail AND (bank$\_$account $=$ enough OR bank$\_$account $=$ not$\_$enough)
\end{itemize}

\section{Exercise \ref{Exercise:Bank3}}
\label{Exercise:Bank3:solution}
The following questions can be asked to help clarify the the description in Exercise \ref{Exercise:Bank1}
\begin{itemize}
    \item What happens if the withdrawal is less than $\$20$?
    \item Is there a time when the ATM doesn't access the central DB?
\end{itemize}

\section{Exercise \ref{Exercise:Bank6}}
\label{Exercise:Bank6:solution}
The answer is option two, specifically what happens if the withdrawal amount is equal to $\$20$.

\section{Exercise \ref{Exercise:Cartesian1}}
\label{Exercise:Cartesian1:solution}
The answer is option two.  Three sets of attributes each with three values results in a Cartesian product of nine.

\section{Exercise \ref{Exercise:Cartesian2}}
\label{Exercise:Cartesian2:solution}
The answer is option one.  N sets of attributes each with k values results in a Cartesian product of n * k.

\section{Exercise \ref{Exercise:Cartesian3}}
\label{Exercise:Cartesian3:solution}
The answer is option one, ${(1,a),(2,a),(3,a),(1,b),(2,b),(3,b)}$

\section{Exercise \ref{Exercise:Cartesian4}}
\label{Exercise:Cartesian4:solution}
The answer is option one, ${(1,b),(2,b),(3,b)}$

\section{Exercise \ref{Exercise:Shopping1}}
\label{Exercise:Shopping1:solution}
You can use the following to help choose tests for the Online Shopping System
\begin{enumerate}
    \item Importance of each attribute
    \item Relations between attributes
    \item Good path vs bad path
     \item Build a table
     \item Client/SME feedback
\end{enumerate}

\section{Exercise \ref{Exercise:Shopping2}}
\label{Exercise:Shopping2:solution}
The total number of possible tests are 288.  To answer how many tests you actually need, you should explore the following questions.  
\begin{enumerate}
    \item Do you have the resources to run all the possible tests?
    \item Do you have limited resources and should perform a test reduction?
    \item Are there constraints that could be applied to this Online Shopping System?
\end{enumerate}

\section{Exercise \ref{Exercise:Shopping3}}
\label{Exercise:Shopping3:solution}
Reviewing your choice can be to talk with developers of the Online Shopping System to discuss the pros and cons of the chosen tests.

\section{Exercise \ref{Exercise:Shopping5}}
\label{Exercise:Shopping5:solution}
The 8 combinations for for \textbf{Credit}, \textbf{One Day} and \textbf{Fedex} are as follows:
\begin{enumerate}
    \item Payment=Credit, Delivery=One Day, Carrier=Fedex, Availability=\textbf{Available}, Export Control=\textbf{True}
    \item Payment=Credit, Delivery=One Day, Carrier=Fedex, Availability=\textbf{Available}, Export Control=\textbf{False}
     \item Payment=Credit, Delivery=One Day, Carrier=Fedex, Availability=\textbf{Not in Stock}, Export Control=\textbf{True}
    \item Payment=Credit, Delivery=One Day, Carrier=Fedex, Availability=\textbf{Not in Stock}, Export Control=\textbf{False}
    Control=\textbf{False}
     \item Payment=Credit, Delivery=One Day, Carrier=Fedex, Availability=\textbf{Discontinued}, Export Control=\textbf{True}
    \item Payment=Credit, Delivery=One Day, Carrier=Fedex, Availability=\textbf{Discontinued}, Export Control=\textbf{False}
    \item Payment=Credit, Delivery=One Day, Carrier=Fedex, Availability=\textbf{No Such Product}, Export Control=\textbf{True}
    \item Payment=Credit, Delivery=One Day, Carrier=Fedex, Availability=\textbf{No Such Product}, Export Control=\textbf{False}
    
\end{enumerate}

\section{Exercise \ref{Exercise:Shopping6}}
\label{Exercise:Shopping6:solution}
There are 36 combinations for Delivery schedule=\textbf{One Day} AND Export control=\textbf{True}.  We would get 4 values for Availability times 3 values for Payment times 3 values for Carrier.  $4*3*3 = 24$

\section{Exercise \ref{Exercise:Interaction1}}
\label{Exercise:Interaction1:solution}
The best answer is 4 variables would be needed.

\section{Exercise \ref{Exercise:Interaction2}}
\label{Exercise:Interaction2:solution}
The best answer is 2 variables would be needed.

\section{Exercise \ref{Exercise:Interaction3}}
\label{Exercise:Interaction3:solution}
There are 302 triplets in the level 3 interaction.  We get 302 by adding the following
\begin{itemize}
    \item 4x3x3 for Availability AND Payment AND Carrier
    \item 4x3x4 for Availability AND Payment AND Delivery
    \item 4x3x2 for Availability AND Payment AND Export
    \item 4x3x4 for Availability AND Carrier AND Delivery
    \item 3x3x2 for Availability AND Carrier AND Export
    \item 4x4x2 for Availability AND Delivery AND Export
    \item 3x3x4 for Payment AND Carrier AND Delivery
    \item 3x3x2 for Payment AND Carrier AND Export
    \item 3x4x2 for Payment AND Delivery AND Export
    \item 3x4x2 for Carrier AND Delivery AND Export
    \item 
\end{itemize}

\section{Exercise \ref{Exercise:Interaction4}}
\label{Exercise:Interaction4:solution}
Option 1 (twelve level 2 interactions)
\begin{itemize}
    \item 2x2 pairs for X and Y 
    \begin{itemize}
        \item x=a,y=a
        \item x=a,y=b
        \item x=b,y=a
        \item x=b,y=b
    \end{itemize}
    \item 2x2 pairs for X and Z
    \begin{itemize}
        \item x=a,z=a
        \item x=a,z=b
        \item x=b,z=a
        \item x=b,z=b
    \end{itemize}
    \item 2x2 pairs for Y and Z
    \begin{itemize}
        \item y=a,z=a
        \item y=a,z=b
        \item y=b,z=a
        \item y=b,z=b
    \end{itemize}
\end{itemize}

\section{Exercise \ref{Exercise:Interaction5}}
\label{Exercise:Interaction5:solution}
Option 2 (eight interaction level two)
\begin{itemize}
    \item 1x2 pairs for X and Y 
    \begin{itemize}
        \item x=b,y=c
        \item x=b,y=d
    \end{itemize}
    \item 1x2 pairs for X and Z
    \begin{itemize}
        \item x=b,z=e
        \item x=b,z=f
    \end{itemize}
    \item 2x2 pairs for Y and Z
    \begin{itemize}
        \item y=c,z=e
        \item y=c,z=f
        \item y=d,z=e
        \item y=d,z=f
    \end{itemize}
\end{itemize}

\section{Exercise \ref{Exercise:Interaction6}}
\label{Exercise:Interaction6:solution}
The model can test for all interaction level 2 but still miss a but on interaction level two of the given model if the bug is really an interaction level 3 type bug and that level 3 interaction isn't covered in the level 2 model.

\section{Exercise \ref{Exercise:Interaction7}}
\label{Exercise:Interaction7:solution}
There are 10 pairs that are covered in this exercise.  They are as follows:
\begin{itemize}
    \item Availability=available, Payment=paypal
    \item Availability=available, Carrier=fedEx
    \item Availability=available, Schedule=$2-5$working days
    \item Availability=available, ExportControl=True
    \item Payment=paypal, Carrier=fedEx
    \item Payment=paypal, Schedule=$2-5$working days
    \item Payment=paypal, ExportControl=True
    \item Carrier=fedEx, Schedule=$2-5$working days
    \item Carrier=fedEx, ExportControl=True
    \item Schedule=$2-5$working days, ExportControl=True
\end{itemize}
Without counting, you can get the number of pairs by multiplying the number of attributes (5) times the coverage we are looking at (pairs=2).  $5*2=10$

\section{Exercise \ref{Exercise:Interaction8}}
\label{Exercise:Interaction8:solution}
Yes, there is a difference. For example, interaction between availability and delivery schedule is not required to be covered by the second option.

\section{Exercise \ref{Exercise:Interaction9}}
\label{Exercise:Interaction9:solution}
The following six pairs are guaranteed by the first coverage requirement but not the second:
\begin{itemize}
    \item Availability, Carrier
    \item Availability, Delivery Schedule
    \item Availability, Export Control
    \item Payment, Carrier
    \item Payment, Delivery Schedule
    \item Payment, Export Control
\end{itemize}

\section{Exercise \ref{Exercise:AddingAttributes}}
\label{Exercise:AddingAttributes:solution}
The best answer is Attribute1 or Attribute2

\section{Exercise \ref{Exercise:AddingAttributes2}}
\label{Exercise:AddingAttributes2:solution}
The best answer is No Attribute

\section{Exercise \ref{Exercise:FunctionCall}}
\label{Exercise:FunctionCall:solution}
The complete Cartesian product is $10 * 2 * 4 = 80$ tests

\section{Exercise \ref{Exercise:manualCTD}}
\label{Exercise:manualCTD:solution}
\begin{itemize}
    \item (red,small,1); (red,medium,2); (red,large,3)
    \item (white,small,2); (white,medium,3); (white,large,1)
    \item (blue,small,3); (blue,medium,1); (blue,large,2)
\end{itemize}

\section{Exercise \ref{Exercise:pairwiseNum}}
\label{Exercise:pairwiseNum:solution}
\begin{enumerate}
    \item 256
    \item 7
\end{enumerate}

\end{document}